\documentclass{IEEEtran}

\usepackage{ifpdf}
\usepackage{cite}
\ifCLASSINFOpdf
\usepackage[pdftex]{graphicx}
\DeclareGraphicsExtensions{.pdf,.jpeg,.png}
\else
\usepackage[dvips]{graphicx}
\DeclareGraphicsExtensions{.pdf}
\fi
\graphicspath{{./Figures/}}
\usepackage{graphicx}
\ifCLASSOPTIONcompsoc
\usepackage[caption=false, font=normalsize, labelfont=sf, textfont=sf]{subfig}
\else
\usepackage[caption=false, font=footnotesize]{subfig}
\fi
\usepackage{balance}
\usepackage{epstopdf}
\usepackage{amsmath,amssymb,mathtools}
\DeclareMathAlphabet{\mathpzc}{OT1}{pzc}{m}{it}
\usepackage{color}
\usepackage{float}
\usepackage[ruled, vlined, linesnumbered]{algorithm2e}
\usepackage{upgreek}
\usepackage{graphicx}

\usepackage{array}
\usepackage{filecontents} 
\usepackage{multirow} 
\usepackage{booktabs} 
\usepackage[mathcal]{euscript} 
\usepackage{tabularx} 
\usepackage{url}
\usepackage{amsthm,bbm}
\usepackage{tabularx}
\usepackage{multirow}
\usepackage{booktabs} 
\usepackage{makecell} 
\usepackage{aligned-overset} 
\usepackage{lipsum} 
\usepackage{bbold}

\newtheorem{Lemma}{Lemma}
\newtheorem{Theorem}{Theorem}
\newtheorem{Corollary}{Corollary}

\newcommand{\uav}{{\sf U}}
\newcommand{\centeruser}{{\sf C}}
\newcommand{\edgeuser}{{\sf E}}
\newcommand{\fusioncenter}{{\sf F}}

\newcommand{\hCF}{ {h_{\centeruser\fusioncenter}} }

\newcommand{\pl} { {\rm PL} }

\newcommand{\area}{ \text{A} }
\newcommand{\threshold}{ {\sf th} }
\newcommand{\thresholdxC}{ {R_{\sf th, C}} }

\newcommand{\loss}{{\pl}}
\newcommand{\los}{\text{LoS}}
\newcommand{\nlos}{\text{NLoS}}

\newcommand{\UEC}{\text{UE-C}}
\newcommand{\UEE}{\text{UE-E}}

\newcommand{\event}{\text{E}}
\newcommand{\success}{\text{S}}
\newcommand{\correct}{\text{C}}

\newcommand{\ibf}[1]{\boldsymbol{\mathit{#1}}}
\renewcommand{\vec}[1]{\boldsymbol{\mathrm{#1}}}

\begin{document}
\setlength\arraycolsep{1pt}
\setlength{\abovedisplayskip}{5pt}
\setlength{\belowdisplayskip}{5pt}

\title{Adaptive Decoding Mechanisms for \\ UAV-enabled Double-Uplink Coordinated NOMA}

\author{ 
	Thanh Luan~Nguyen, Georges~Kaddoum,~\IEEEmembership{Senior~Member,~IEEE,}~Tri~Nhu~Do, Daniel~Benevides~da~Costa,~\IEEEmembership{Senior~Member,~IEEE}~and~Zygmunt~J.~Haas,~\IEEEmembership{Fellow,~IEEE}
    
	\thanks{T.-L. Nguyen, G.~Kaddoum and T.~N.~Do are with the Department of Electrical Engineering, the \'{E}cole de Technologie Sup\'{e}rieure (\'{E}TS), Universit\'{e} du Qu\'{e}bec, Montr\'{e}al, QC H3C 1K3, Canada. {G.~Kaddoum is also with Cyber Security Systems and Applied AI Research Center, Lebanese American University} (emails: thanh-luan.nguyen.1@ens.etsmtl.ca, georges.kaddoum@etsmtl.ca, tri-nhu.do@etsmtl.ca).}

	\thanks{D. B. da Costa is with the Technology Innovation Institute, 9639 Masdar City, Abu Dhabi, United Arab Emirates (email: danielbcosta@ieee.org).}
	
	\thanks{Z. J. Haas is with the Department of Computer Science, University of Texas at Dallas, Richardson, TX 75080, USA, and also with the School of Electrical and Computer Engineering, Cornell University, Ithaca, NY 14853, USA (e-mail: zhaas@cornell.edu).}
	
}

\maketitle

\begin{abstract}
In this paper, we propose a novel adaptive decoding mechanism (ADM) for the unmanned aerial vehicle (UAV)-enabled uplink (UL) non-orthogonal multiple access (NOMA) communications. 
    Specifically, considering a harsh UAV environment, where ground-to-ground links are regularly unavailable, the proposed ADM overcomes the challenging problem of conventional UL-NOMA systems whose performance is sensitive to the transmitter's statistical channel state information and the receiver's decoding order. 
To evaluate the performance of the ADM, we derive closed-form expressions for the system outage probability (OP) and system throughput. 
    In the performance analysis section, we provide novel expressions for practical air-to-ground and ground-to-air channels, while taking into account the practical implementation of imperfect successive interference cancellation (SIC) in UL-NOMA. 
    Moreover, the obtained expression can be adopted to characterize the OP of various systems under a Mixture of Gamma (MG) distribution-based fading channels. 
Next, we propose a sub-optimal Gradient Descent-based algorithm to obtain the power allocation coefficients that result in maximum throughput with respect to each location on UAV's trajectory. 
    To determine the significance of the proposed ADM in nonstationary environments, we consider the ground users and the UAV to move according to the Random Waypoint Mobility (RWM) and Reference Point Group Mobility (RPGM) models, respectively. Accurate formulas for the distance distributions are also provided.
Numerical solutions demonstrate that the ADM-enhanced NOMA not only outperforms Orthogonal Multiple Access (OMA), but also improves the performance of UAV-enabled UL-NOMA even in mobile environments.
\end{abstract}


\section{Introduction}

Due to its potential for wide-area coverage, scalability, service continuity, and improved availability \cite{GiordaniIN2021}, the development of non-terrestrial networks (NTNs) is attracting significant research interests, fueled by the introduction of the beyond-fifth-generation (B5G) and the sixth generation (6G) wireless networks. NTNs include low-altitude platforms (LAPs), which have been proposed for various military and civilian mission-critical applications \cite{CaoJSAC2018}. Among key components of LAPs are unmanned aerial vehicles (UAVs), which are flexible, low cost, and close to the
ground platforms.
    As a result, a variety of UAV-based technologies and applications have been proposed. These include physical-layer security and emergency communications in post-disaster scenarios or in unexpected events \cite{MozaffariCST2019, WangMVT2021}.
    UAVs can also act as aerial relays in cooperative wireless communications to assist users with poor channel conditions, such as at cell-edge locations in cellular networks \cite{WuJSAC2018}. 
The reason for this is that UAVs can change their position to improve channel conditions, allowing the performance of cooperative transmissions to be greatly improved even when direct communication between the transmitter and the receiver is unavailable.
    
In the context of NTNs, Non-Orthogonal Multiple Access (NOMA) plays a key role in future B5G and 6G wireless networks networks, which effectively exploits available resources such as frequency, time, power, and code, as to ensure multiple users meet their quality-of-service (QoS) requirements \cite{DaiCST2018, SerghiouCST2022, VaeziCST2022}.
     Specifically, NOMA was shown to be a promising technique and has attracted considerable attention from both academia and industry \cite{BarickAccess2022, KatweTWC2022, HuIoT2022}. 
In NOMA, the receivers recover information received from paired transmitters, by employing successive interference cancellation (SIC) with a designed decoding order \cite{ClerckxOJCS2021, LiuJSAC2022}.
    As a result, the use of NOMA provides additional degrees of freedom in the power domain and enhancement of spectrum utilization.
This promising technique has been intensively investigated in the literature in various systems, such as coordinated direct and relay transmission (CDRT-NOMA) \cite{NguyenLWC2021, PanLWC2021}, massive multiple-input-multiple-output (mMIMO-NOMA) \cite{JawarnehAccess2022}, secure systems \cite{LvTVT2022}, and reconfigurable intelligent surface networks (IRS-NOMA) \cite{LiTGCN2022}.

In the aforementioned studies, limited efforts have been made to integrate UAVs with CDRT-NOMA systems. 
    Using UAVs as intermediate relays in CDRT-NOMA systems can increase the coverage area, extend the communication range of cell-edge users ($\UEE$s), and provide better spectral efficiency (SE).
However, the dynamic motion of UAVs remains a key challenge, as unpredictable fluctuations in channel information and signal quality can significantly impair communication range and overall system performance.
    For downlink CDRT-NOMA, the authors in \cite{NguyenLWC2021} presented an CDRT-NOMA scheme for IoT networks seeking to improve spectrum efficiency (SE). The authors provided results for outage probability (OP) and Ergodic Sum Capacity (ESC).
Moreover, the study in \cite{XuTC2021} proposed spectrum-efficient schemes for CDRT-NOMA under the assumption of Nakagami-$m$ fading channels.
    However, studies on uplink (UL) scenarios are quite limited, which can prevent CDRT-NOMA systems from being deployed in a variety of
applications such as navigation, disaster relief, and ubiquitous
data collection.

In conventional UL CDRT-NOMA systems, there are two information transmission phases \cite{PanLWC2021, KaderLCOMM2017, XuAccess2019}. 
    The first phase involves the cell-center user ($\UEC$) and $\UEE$ transmitting information to a fusion center (FC) and an intermediate relay, respectively.
    In the second phase, the $\UEC$ transmits new information, and the relay node forwards the decoded information from $\UEE$, causing interference at the FC.
In both phases, the SIC receivers at the relay and at the FC exploit the distinct channel gain of from $\UEC$ and $\UEE$ to correctly decode the desired information signals.
    
    However, when UAVs are treated as an aerial relay, the unpredictable motion along predefined flight paths poses a challenge for the SIC architecture, especially the fixed SIC architecture, which could degrade the system's performance.
    
    The study in \cite{KaderWCL2018} proposed an UL CDRT-NOMA scheme and derived the ESC under both perfect and imperfect SIC at the receivers, showing a significant capacity improvement compared to traditional orthogonal multiple access (OMA) schemes.
In \cite{PanLWC2021}, the authors proposed relaying schemes that improve the system throughput of two-user CDRT-NOMA when the transmitters have similar channel conditions.
    Moreover, the authors in \cite{XuAccess2019} considered hybrid device-to-device (D2D) and uplink CDRT-NOMA to enhance the SE and extend the coverage area of the cellular network.

Nevertheless, the existing UL CDRT-NOMA schemes cannot treat UAVs as terrestrial relays because of fundamental challenges discussed below.
    In these schemes, the non-adaptive SIC order is based on the statistical channel state information (CSI) of the transmitters, where the receiver exploits a large amount of instantaneous CSI to decode the received signals from the transmitters \cite{LiuCL2018, AgarwalAccess2020, NguyenSJ2022}.
In \cite{LiuCL2018}, based on statistical CSI, the authors derived the OP of multi-user UL NOMA schemes over Rayleigh channels.
    An extension to generalized fading was developed in  \cite{AgarwalAccess2020}.
A recent research on the use of CDRT-NOMA in Internet-of-Things (IoT) networks promoting statistical CSI-based decoding order was reported in  \cite{NguyenSJ2022}. 
    However, due to the high mobility of UAVs, the statistical CSI of the UAV to terrestrial device links fluctuates dynamically as the UAV travels in the three-dimensional (3D) space \cite{SharmaWCL2019, AmerTC2020}.
As a result, the UAV and FC can experience extreme interference due to the non-adaptive decoding orders, leading to incorrect signal segregation at the SIC receivers, which limits the application of such schemes.
    To address this issue, a more adaptive SIC ordering should be considered.

The available decoding mechanisms in \cite{GaoCL2017, GaoTVT2018, ZakeriTVT2019} require not only perfect channel state information (CSI), but also information about the power control strategy. 
    The studies in \cite{GaoCL2017, GaoTVT2018}, in particular, present dynamic decoding orders for uplink NOMA transmission, where dynamic SIC receivers estimate the received power from each transmitter before decoding the received signal. 
In addition, \cite{ZakeriTVT2019} proposed resource allocation and power control algorithms for dynamic SIC receivers.

In this paper, we consider uplink CDRT-NOMA in wireless cellular networks. Specifically, our focus is on a scenario in which the network includes a $\UEC$ that communicates directly with the FC, and a $\UEE$ with no direct access to the FC. 
    To overcome this challenge, $\UEE$ requires the assistance of a decode-and-forward (DF) UAV serving as the relay node to enable communication with the FC. 
However, our DF relaying strategy differs from the conventional DF one. Specifically, the UAV separates the signals from $\UEE$ and $\UEC$ using ADM-assisted successive interference cancellation (SIC), which allows for the extraction of each individual signal even in the presence of dynamic channel disparities. As a result, the DF technique in the proposed system is considerably more complex and sophisticated than the traditional method.
 In contrast to related works, we consider an adaptive UL CDRT-NOMA transmission, where the UAV and the FC adaptively decode the UL signals by exploiting the UL channel power gains.
The contributions of this paper can be summarized as follows:

\begin{itemize}
    \item {We propose novel adaptive decoding mechanisms (ADMs) tailored for Double-UL (DUL) power-domain NOMA (PD-NOMA) systems. 
    Implemented at the unmanned aerial vehicle (UAV) and at the fusion center (FC). These mechanisms enable efficient DUL transmissions, even in the presence of dynamic channel disparities. 
Unlike existing SIC-inherited systems that rely on instantaneous received power, the proposed ADM addresses the need for adaptation to channel fluctuations by relying solely on instantaneous channel state information (CSI), while enabling superior system performance.}
    
    \item Under this setting, a new framework is developed to derive several challenging OP formulas of the proposed system.
    For the sake of tractability, general expressions for successful decoding probabilities under Mixture of Gamma (MG)-based fading channels are derived and used to obtain various closed-expressions of the OPs.
    
    \item To improve the cell-edge performance, we formulate an optimal power allocation strategy that minimizes the OP of the edge user. To this end, a fast-convergent Gradient Decent algorithm is proposed to achieve a near-optimal solution with a negligible optimality gap\footnote{It is noted that our results are reproducible using our Matlab code, which is available at \url{https://github.com/thanhluannguyen/UAV-NOMA-ADM}.}.

    \item Then, we evaluate the performance of the proposed ADM in nonstationary conditions, where UEs and UAVs are mobile, and move according to Random Waypoint Mobility (RWM) and Reference Point Group Mobility (RPGM) models{\cite{Hong1999, HyytiaTMC2006, mobility}${}^{\text{1}}$}. In addition, a variety of distance distributions between mobile nodes are derived in closed-form expressions with high analytical precision.

    \item We compare the proposed ADM and all possible Non-Adaptive Decoding Mechanisms (NADMs) for the DUL CDRT-NOMA under nonstationary conditions. Closed-form expressions for the average throughput are provided to examine the impact of mobile nodes' velocity. The results show that ADM can provide stable throughput even when both UEs and UAV have high movement~velocity.
\end{itemize}

\noindent \textbf{Notations}: $\mathbb{E} [\cdot]$ denotes the expectation operator; $f_X (x)$, $F_X (x)$, and $F^c_X(x)$ represent the probability density function (PDF), cumulative distribution function (CDF), and the complementary CDF (CCDF) of an arbitrary random variable (RV) $X$, respectively; $\Pr[\mathrm{A}]$ stands for the probability that an event $A$ occurs; $\mathcal{CN} (0,\sigma^2)$ denotes a circular symmetric complex Gaussian RV with zero-mean and variance $\sigma^2$; and $\Gamma(\cdot,\cdot)$ and $\gamma(\cdot,\cdot)$ denote the upper and lower incomplete gamma functions \cite[Eq. (8.350.1)]{Gradshteyn2007}, respectively.  

\section{System Model}

\begin{figure}[t]
    \centering
    \includegraphics[width=\linewidth]{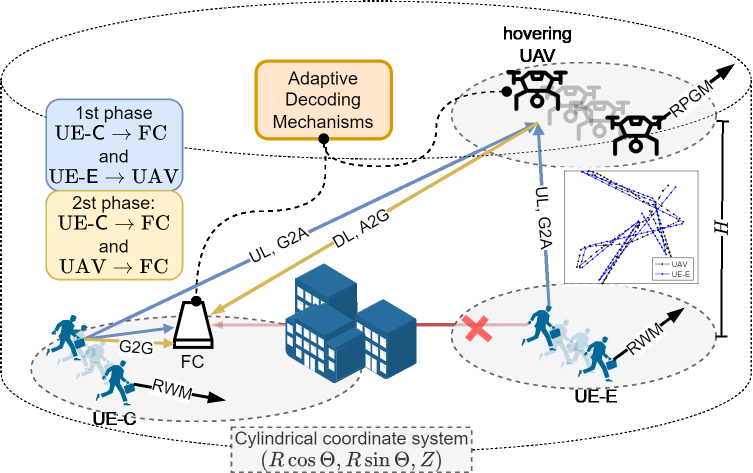}
    \caption{The proposed UAV-aided DUL coordinated NOMA with ADM, where $\UEE$ and $\UEC$ move in accordance with the RPGM model and UAV follows $\UEE$ in accordance with the RPGM model.}
    \label{fig:sysModel}
\vspace{-0.3cm}
\end{figure}

Let us consider DUL transmissions of an IoT system consisting an $\UEC$ node and an $\UEE$ node, communicating with a FC node, which can be a wireless access point, as depicted in Fig. \ref{fig:sysModel}. 
    Hereafter, let us use $\centeruser$, $\edgeuser$, $\fusioncenter$ and $\uav$ as acronyms to represent $\UEC$, $\UEE$, FC, and UAV, respectively.
Due to radio propagation impairments, the link between $\edgeuser$ and $\fusioncenter$ is unavailable, thus the UL transmission of $\edgeuser$ relies on the assistance of an intermediate $\uav$, acting as a relay.
    We first assume that the locations of the nodes are fixed during the whole communication \cite{HouraniWCL2014, MeiJSTSP2019, YuanTVT2022}.
We adopt the three-dimensional (3D) Cartesian coordinates system to model the locations of $\centeruser$, $\edgeuser$, $\fusioncenter$, and $\uav$ as $(X_\centeruser, Y_\centeruser, Z_\centeruser)$, $(X_\edgeuser, Y_\edgeuser, Z_\edgeuser)$, $(X_\fusioncenter, Y_\fusioncenter, Z_\fusioncenter)$ and $(X_\uav, Y_\uav, Z_\uav)$, respectively.
The equivalent coordinates of $\uav$ in the 3D cylindrical coordinates system are given as $(R\cos\Theta, R\sin\Theta, Z_\uav)$, where $R$ and $\Theta$ are the radius and angular coordinates of $\uav$, respectively.
The distance between ${\sf X}$ and ${\sf Y}$, denoted as $d_{\sf XY} $, can be formulated as
\begin{align}
d_{\sf XY} 
    \!=\! \sqrt{ (X_{\sf Y}-X_{\sf X})^2 \!+\! (Y_{\sf Y}-Y_{\sf X})^2 \!+\! (Z_{\sf Y}-Z_{\sf X})^2 }, 
\end{align}
where ${\sf XY} \in \{ {\centeruser\uav}, {\uav\fusioncenter}, {\centeruser\fusioncenter}, {\edgeuser\uav} \}$.
We denote $h_{\sf XY}$ as the channel coefficient between the nodes ${\sf X}$ and ${\sf Y}$, and assume that all channels are reciprocal.
    The channel coefficient $h_{\sf XY}$ can be expressed as
\begin{align}
h_{\sf XY}
    =   g_{\sf XY} e^{j \angle h_{\sf XY}}  \sqrt{{\pl}_{\sf XY} },
\label{eq:channel_coeff}
\end{align}
where $g_{\sf XY}$ and $\angle h_{\sf XY}$ are the magnitude and phase of the small-scale fading, respectively, and $\sqrt{ {{\pl}_{\sf XY}} }$ is the large-scale fading.
In addition, the DUL transmission scheme can be divided into two phases as follows. During the first phase, of duration $t_1$, $\centeruser$ and $\edgeuser$ transmit their information; i.e., $x^{[1]}_{\centeruser}$ and $x_{\edgeuser}$, to $\fusioncenter$ and $\uav$, respectively. 
During the second phase of duration $t_2$, $\centeruser$ and $\uav$ convey $x^{[2]}_{\centeruser}$ and $\hat{x}_{\edgeuser}$ to $\fusioncenter$, respectively. 
 
\subsection{Ground-to-Ground (G2G) Channel Model}

We assume that $\centeruser$ is located within some distance of $\fusioncenter$.
    In other words, the line-of-sight (LoS) links between $\centeruser$ and $\fusioncenter$ can be impacted by shadowing, such as being obscured by buildings.
In this paper, we assume the channel between $\centeruser$ and $\fusioncenter$ follows the shadowed Rician fading model, where the probability density function (PDF) of $g_{\centeruser\fusioncenter}^2$ is obtained as \cite{EspinosaGLOBECOM2019}
\begin{align}
f_{g_{\centeruser\fusioncenter}^2}(x)
    =   \alpha 
        \!\!\!\sum_{l=1}^{m_{\centeruser\fusioncenter}-1}
        \zeta(l) \frac{ x^{l} }{ l! } e^{-(\beta-\delta) x}
    +   \alpha e^{ -(\beta-\delta) x },~ x > 0, 
\end{align}
where ${\zeta(l) \triangleq \frac{ (-1)^l \langle 1-m_{\centeruser\fusioncenter}\rangle_l \delta^l }{ l! }}$ with $\langle \cdot\rangle_l$ being the Pochhammer symbol \cite{Gradshteyn2007},
${m_{\centeruser\fusioncenter}}$ is the shadowing severity parameter,
${\beta \!\triangleq\! \frac{1}{2 b}}$,
${\alpha \triangleq 
\frac{1}{2b}
\left( 
    \frac{2 b m_{\centeruser\fusioncenter}}{2 b m_{\centeruser\fusioncenter} + \Omega_{\centeruser\fusioncenter} }
\right)^{ m_{\centeruser\fusioncenter} } }$, and
${\delta \triangleq \frac{\Omega_{\centeruser\fusioncenter}}{ 2b(2 b m_{\centeruser\fusioncenter} + \Omega_{\centeruser\fusioncenter} ) }}$ with 
${\Omega_{\centeruser\fusioncenter}}$ and $2 b$ being the average power of the LoS and multipath components, respectively.
    For the large-scale fading, we consider the path loss model ${ {\pl}_{\centeruser\fusioncenter}(d_{\centeruser\fusioncenter}) \!\triangleq\! G_{\centeruser}~[\text{dB}] + G_{\fusioncenter}~[\text{dB}] - 10\log{\pl} - 10\epsilon\log(\frac{d_{\centeruser\fusioncenter}}{d_0}) }$ [dB], where $G_{\centeruser}$ and     $G_{\fusioncenter}$ are the transmit and receive antenna gains at $\centeruser$ and $\fusioncenter$, respectively,
    $d_0 = 1$ [m] is the reference distance, 
    $\epsilon$ is the path loss exponent, 
    and ${\pl}$ is the average power attenuation factor
measured at the carrier frequency $f_c$ and at $d_0$. 
    {From \eqref{eq:channel_coeff}, the resulted G2G channel power gain can be expressed as $|h_{\centeruser\fusioncenter}|^2 = g^2_{\centeruser\fusioncenter} {\rm PL}_{\centeruser\fusioncenter}$. }

\vspace{-0.2cm}
\subsection{Air-to-Ground (A2G) and Ground-to-Air (G2A) Channel Models}

In our model, there may be limited information on the precise locations, heights, and amount of radio impairments, such as due to buildings, walls, and so on.
    Meanwhile, the randomness associated with the LoS and non-line-of-sight (NLoS) links must be considered in the design of the UAV-based communication system. 
As a result, to develop practical A2G/G2A channels, we consider that each device, i.e., $\edgeuser$, $\centeruser$, and $\fusioncenter$, has a probabilistic LoS link towards $\uav$.
    The LoS probability is a function of the device's and $\uav$'s locations in the 3D plane.
One popular approach for determining the LoS probability is based on a logistic function, which is expressed as
$p^{\rm LoS}_{\sf XY}
    = ( 1+ \alpha_1 e^{ -\alpha_2(J_{\sf XY}-\alpha_1)} )^{-1}$,
where $\alpha_1$ and $\alpha_2$ are environmental-dependent coefficients \cite{HouraniWCL2014, MozaffariCST2019, YuanTVT2022}, 
$J_{\sf XY}$ [deg] denotes the elevation angle  \cite{HouraniWCL2014}.
    Specifically, $J_{\sf XY}$ can be calculated as
$
J_{\sf XY}
    \!=\! \frac{180}{\pi} 
    \arcsin 
        \frac{ \left| Z_{\sf Y}-Z_{\sf X} \right|}
        {d_{\sf XY}}
$,
where $|\!\cdot\!|$ denotes the absolute value operator. In addition, the probability of having a NLoS connection is $p^{\rm NLoS}_{\sf XY} = 1 - p^{\rm LoS}_{\sf XY}$.
The PDF of $g_{\sf XY}$ for the A2G/G2A channels can be modeled as \cite{HouraniWCL2014, MozaffariCST2019, YuanTVT2022}
\begin{align}
f_{g_{\sf XY}}(x;m)
    =   \left\{
    \begin{array}{cc}
    \displaystyle
        \frac{2 m^m}{(m-1)!} x^{2 m-1} e^{-m x^2},
            \hspace{5pt} \text{for LoS link}, \vspace{3pt} \\
    \displaystyle
        2x e^{-x^2},
            \hspace{5pt} \text{for NLoS link},
    \end{array}
    \right.
\end{align}
for $x > 0$, where $m$ is the shape factor, which corresponds to the Nakagami-$m$ fading for LoS link.

The path loss of the A2G/G2A channels takes into account probabilistic LoS and NLoS conditions. When the UAV is hovering, the path loss for LoS and NLoS links are modeled as \cite{HouraniWCL2014, MozaffariCST2019, YuanTVT2022}
\begin{align}
{\pl}_{\sf XY}~[\text{dB}]
    = \left\{
    \begin{array}{ll}
        {\pl}^{\los}_{\sf XY},~\text{for LoS link}, \vspace{5pt} \\
        {\pl}^{\nlos}_{\sf XY},~\text{for NLoS link},
    \end{array}\!\!
    \right.\!
\end{align}
where 
\begin{subequations}
\begin{align}
{\pl}^{\los}_{\sf XY} &=\! 
    G_{\sf X} + G_{\sf Y} +
    20\log\left(
        \frac{ \frac{c}{f_c} }{ 4\pi d_{\sf XY} }
    \right) \!-\! \eta~[\text{dB}], \\
{\pl}^{\nlos}_{\sf XY} &=\! 
    G_{\sf X} + G_{\sf Y} +
    20\log\left(
        \frac{ \frac{c}{f_c} }{ 4\pi d_{\sf XY} }
    \right) \!-\! \bar{\eta}~[\text{dB}],
\end{align}
\end{subequations}
where $G_{\sf X}$ and $G_{\sf Y}$ are transmit and receive antenna gains, respectively, 
    $c$ is the speed of light, $\eta$ and $\bar{\eta}$ are the link attenuation under LoS and NLoS links, respectively.
Given $\iota_{\sf XY}$ as an identity RV representing the LoS condition of the A2G links, where $\mathbb{E}[\iota_{\sf XY}] = p^{\los}_{\sf XY}$, ${\sf XY} \in \{ \uav\fusioncenter, \edgeuser\uav, \centeruser\uav \}$, the resulting A2G channel power gain can be expressed as 
$   |h_{\sf XY}|^2 
        = \iota_{\sf XY} g^2_{\sf XY}|_{\los} {\rm PL}_{\sf XY}^{\los}
        + (1-\iota_{\sf XY}) g^2_{\sf XY}|_{\nlos} {\rm PL}^{\nlos}_{\sf XY}$.

\vspace{-0.2cm}
\subsection{The Proposed ADM}

	Note that with the proposed DUL transmission, the performance of $\centeruser$ is not compromised by time-division in Decode-and-Forward (DF) relaying as is used in conventional relaying schemes, and the severe fading conditions experienced by $\edgeuser$ can be remedied, guaranteeing  a reliable communication. 
    Such an application of this scheme is employed when the delay-limited service-based $\centeruser$ requires high data rate (e.g., video streaming services), whereas the delay-tolerant service-based $\edgeuser$ demands a reliable communication.
    
\subsubsection{The First UL Phase}

Following the settings of PD-NOMA-aided uplink transmission \cite{MeiJSTSP2019} during the first phase, $\edgeuser$ and $\centeruser$ simultaneously transmit $x_{\edgeuser}$ and $x^{[1]}_{\centeruser}$ to $\uav$ and $\fusioncenter$, respectively.
 The transmission powers of $\edgeuser$ and $\centeruser$ are denoted by $P^{[1]}_{\centeruser}$ and $P_{\edgeuser}$, respectively, where $P^{[1]}_{\centeruser} + P_{\edgeuser} \le P^{[1]}_{\max}$ with $P^{[1]}_{\max}$ being the maximum power budget in the first phase. 
Hence, the received signals at $\uav$ and $\fusioncenter$ are, respectively, formulated as
\begin{align}
y_{\uav}
    &= {h}_{\centeruser\uav} x^{[1]}_{\centeruser} 
    +  {h}_{\edgeuser\uav} x_{\edgeuser}
	+ n_{\uav},
\label{eq:1} \\
y_{\fusioncenter}^{[1]}
    &= \hCF x^{[1]}_{\centeruser} + n_{\fusioncenter}^{[1]},
\label{eq:2}
\end{align}
where
    ${\mathbb{E}\big[
        \big|x^{[1]}_{\centeruser}\big|^2 
    \big] = P^{[1]}_{\centeruser}}$, 
    ${\mathbb{E}\big[ 
        \big|x_{\edgeuser}\big|^2 
    \big] = P_{\edgeuser}}$, 
    ${n_{\uav} \!\sim\! {\cal CN}\left(0,\sigma^{2}_{\uav}\right)}$, and ${n_{\fusioncenter}^{[1]} \sim {\cal CN}\left(0,\sigma^{2}_{\fusioncenter} \right)}$ are the zero means additive white Gaussian noises (AWGNs) at $\uav$ and $\fusioncenter$, respectively, with $\sigma^{2}_{\uav}$ and $\sigma^{2}_{\fusioncenter}$ being the corresponding noise powers.
The achievable rate at $\fusioncenter$ to decode $x^{[1]}_{\centeruser}$ is given by
\begin{equation}
{R}^{[1]}_{\centeruser}
=  \frac{1}{2}
\log_2\bigg(	
	1+\frac{ P^{[1]}_{\centeruser} g_{\sf CF}^2 \loss_{\centeruser\fusioncenter} }{ \sigma^2_{\fusioncenter} } 
\bigg).\!\!
\label{eq_rate_F_xC2}
\end{equation}
Meanwhile, by exploiting the disparity in the gains $g_{\centeruser\uav}^2 {\pl}_{\centeruser\uav}$ and $g_{\edgeuser\uav}^2 {\pl}_{\edgeuser\uav}$, $\uav$ determines an effective decoding mechanism to guarantee reliable transmission at $\edgeuser$.
    Specifically, the proposed ADM at $\uav$ are described next.

In the event
$\{ {\event}^{[\uav]}_{\centeruser\to\edgeuser}
    \!:=\! g^2_{\centeruser\uav} {\pl}_{\centeruser\uav}
        \!\ge\! g^2_{\edgeuser\uav} {\pl}_{\edgeuser\uav} \}$, 
the~channel from $\uav$ to $\centeruser$ is stronger than that from $\uav$ to $\edgeuser$, thus $\uav$ primarily decodes $x^{[1]}_{\centeruser}$ before canceling it from $y_{\uav}$ with imperfect SIC.
    Then, the desired signal for $\uav$, $x_{\edgeuser}$, is decoded. 
The achievable rate at $\uav$ to decode $x^{[1]}_{\centeruser}$ is given by
\begin{equation}
R^{[\uav]}_{\centeruser\to\edgeuser}
    =  \frac{1}{2}
    \log_2\bigg(	
    	1 + \frac{ P_\centeruser g_{\centeruser\uav}^2 {\pl}_{\centeruser\uav} }
        { P_{\edgeuser} g_{\edgeuser\uav}^2 {\pl}_{\edgeuser\uav}
        + \sigma_\uav^2 }
    \bigg).
\label{eq_R_ADMUI_C_xC}
\end{equation}
    In practice, $\uav$ cannot perfectly cancel $x^{[1]}_{\centeruser}$ from $y_{\uav}$ with SIC, which incurs an amount of residual power from $x^{[1]}_{\centeruser}$ and causes additional interference while $\uav$ decodes $x_{\edgeuser}$. 
Subsequently, $\uav$ decodes $x_{\edgeuser}$ with the instantaneous achievable rate given by
\begin{equation}
R^{[\uav]}_{\edgeuser}
= \frac{1}{2}
\log_2\left(
	1 + \frac{ P_{\edgeuser} g_{\edgeuser\uav}^2 {\pl}_{\edgeuser\uav}}{ P^{[1]}_{\centeruser} \big| \tilde{h}_{\centeruser\uav} \big|^2 + \sigma^2_{\uav} }
\right),
\label{eq_R_ADMUI_C_xE}
\end{equation}
where the $P^{[1]}_{\centeruser} \big| \tilde{h}_{\centeruser\uav} \big|^2$ specifies the interference power from imperfectly removing $x^{[1]}_{\centeruser}$, $\tilde{h}_{\centeruser\uav}$ has a circularly-symmetric Gaussian distribution with zero mean and variance $\xi_{\uav} {\pl}_{\centeruser\uav} (d_{\centeruser\uav})$, denoted as $\tilde{h}_{\centeruser\uav} \!\sim\! \mathcal{CN}(0,\xi_{\uav} {\pl}_{\centeruser\uav} (d_{\centeruser\uav}))$, where $\xi_{\uav} \!\in\! [0,1]$ denotes the residual interference (RI) level due to imperfect SIC \cite{KaderLCOMM2017}. 
    It is noteworthy that $\tilde{h}_{\centeruser\uav}$ and ${h}_{\centeruser\uav}$ are independent RVs and $\xi_{\uav} = 0$ represents perfect SIC.

In the event 
    $\{ {\event}^{[\uav]}_{\edgeuser\to\centeruser}
    \!:=\! g^2_{\centeruser\uav} {\pl}_{\centeruser\uav}
        \!<\! g^2_{\edgeuser\uav} {\pl}_{\edgeuser\uav} \}$, which~is~the complementary event of ${\event}^{[\uav]}_{\centeruser\to\edgeuser}$,
	$\uav$ attempts to decode $x_{\edgeuser}$ while treating $x^{[1]}_{\centeruser}$ as interference. 
Since the main objective of cooperative communication is to forward $\edgeuser$'s information signal to $\fusioncenter$, we only consider the decoding of $x_{\edgeuser}$ and neglect $x^{[1]}_{\centeruser}$.
	Subsequently, $\uav$ decodes $x_{\edgeuser}$ with the achievable rate given~by
\begin{equation}
R^{[\uav]}_{\edgeuser\to\centeruser}
=   \frac{1}{2}
\log_2\bigg(
	1 + \frac{ P_{\edgeuser} g_{\edgeuser\uav}^2 {\pl}_{\edgeuser\uav} }{ P^{[1]}_{\centeruser} g_{\centeruser\uav}^2 {\pl}_{\centeruser\uav} + \sigma^2_{\uav} }
\bigg).
\label{eq_rate_ADMU2_U_xE}
\end{equation}

For $R_{\threshold,\centeruser}$ [bits/s/Hz] and $R_{\threshold,\edgeuser}$ [bits/s/Hz] being the minimum target spectral efficiency of $\centeruser$ and $\edgeuser$, respectively,
the UAV's decisions {during the second phase} and the corresponding received SINR events are presented in Table \ref{tab:deci_UAV}, where  
    ${{\success}_{\centeruser\to\edgeuser}^{[\uav]}
    \!:=\! \{ R^{[\uav]}_{\centeruser\to\edgeuser} \!>\! R_{\threshold,\centeruser} \}}$,
    ${{\success}_{\edgeuser}^{[\uav]}
    \!:=\! \{ R^{[\uav]}_{\edgeuser} \!>\! R_{\threshold,\edgeuser} \}}$,
    and
    ${{\success}_{\edgeuser\to\centeruser}^{[\uav]} 
    \!:=\! \{ R^{[\uav]}_{\edgeuser\to\centeruser} \!>\! R_{\threshold,\edgeuser} \}}$ are the decoding events at $\uav$.
{Specifically, the {\it silent} decision indicates that $\uav$ remains inactive, while the {\it forward} decision implies that $\uav$ relays information from $\edgeuser$ to $\fusioncenter$.}

{\renewcommand{\arraystretch}{1.5}
\begin{table}[!t]
\centering
\caption{Decision at the UAV.\label{tab:deci_UAV}}
\begin{tabular}{|c|c|}
\hline
UAV's Decision & SINR Events
\\\hline
silent & $\{ \event_{\centeruser\to\edgeuser}^{[\uav]}, \bar{\success}_{\centeruser\to\edgeuser}^{[\uav]} \}$, $\{ \event_{\centeruser\to\edgeuser}^{[\uav]}, {\success}_{\centeruser\to\edgeuser}^{[\uav]}, \bar{\success}_{\edgeuser}^{\uav} \}$, $\{ \event_{\edgeuser\to\centeruser}^{[\uav]}, \bar{\success}_{\edgeuser\to\centeruser}^{[\uav]} \}$ 
\\\hline
forward & $\{ \event_{\centeruser\to\edgeuser}^{[\uav]}, {\success}_{\centeruser\to\edgeuser}^{[\uav]}, {\success}_{\edgeuser}^{\uav} \}$, $\{ \event_{\edgeuser\to\centeruser}^{[\uav]}, {\success}_{\edgeuser\to\centeruser}^{[\uav]} \}$
\\\hline
\end{tabular}
\vspace{-0.3cm}
\end{table}}

\subsubsection{The Second Uplink Phase}

Assuming $\uav$ correctly decodes $x_{\edgeuser}$ in the first phase, it forwards $\hat{x}_{\edgeuser}$ in the second phase. 
    Meanwhile, $\centeruser$ also transmits $x^{[2]}_{\centeruser}$, which is different from $x^{[1]}_{\centeruser}$. 
In addition, assuming that the transmission powers of $\centeruser$ and $\uav$ in the second phase are $P^{[2]}_{\centeruser}$ and $P_{\uav}$, respectively, where ${P^{[2]}_{\centeruser} \!+\! P_{\uav} \le P^{[2]}_{\max}}$ with $P^{[2]}_{\max}$ being the maximum power budget during this phase, the received signal at $\fusioncenter$ in the case that $x_{\edgeuser}$ was correctly decoded by $\uav$ is expressed as
\begin{equation}
y^{[2]}_{\fusioncenter} = h_{{\uav} \fusioncenter } \hat{x}_{\edgeuser}
	 + \hCF {x}_{\centeruser}^{[2]} + n_{\fusioncenter}^{[2]}.
\end{equation}
Similarly, we also consider ADM at $\fusioncenter$, which are described as follows.
In the scenario that
$\{ {\event}^{[\fusioncenter]}_{\centeruser\to\uav}
    \!:=\! 
    g^2_{\centeruser\fusioncenter} {\pl}_{\centeruser\fusioncenter}
        \!\ge\! g^2_{\uav\fusioncenter} {\pl}_{\uav\fusioncenter} \}$,
the channel from $\fusioncenter$ to $\centeruser$ is stronger than from $\fusioncenter$ to $\uav$, $\fusioncenter$ primarily decodes $x^{[2]}_{\centeruser}$ while treating $\hat{x}_{\edgeuser}$ as interference. 
    In this case, $\fusioncenter$ decodes $x^{[2]}_{\centeruser}$ with the achievable rate given by
\begin{equation}
R^{[\fusioncenter]}_{\centeruser\to\uav}
= \frac{1}{2}
\log_2\bigg(
	1 + \frac{ P^{[2]}_{\centeruser} g_{\centeruser\fusioncenter}^2 {\pl}_{\centeruser\fusioncenter} }{ P_{\uav} g_{\uav\fusioncenter}^2 {\pl}_{\uav\fusioncenter} + \sigma^2_{\fusioncenter} }
\bigg).
\label{eq_R_ADMFI_C_xC}
\end{equation}
Once $x^{[2]}_{\centeruser}$ is correctly decoded, $\fusioncenter$ performs SIC in an attempt to remove $x^{[2]}_{\centeruser}$ from $y^{[2]}_{\fusioncenter}$ before decoding $\hat{x}_{\edgeuser}$. 
    Assuming imperfect SIC, $\fusioncenter$ decodes $\hat{x}_{\edgeuser}$  with the achievable rate given~by

\begin{equation}
R^{[\fusioncenter]}_{\uav}
=   \frac{1}{2}
\log_2\left(
	1 + \frac{ P_{\uav} g_{\uav\fusioncenter}^2 {\pl}_{\uav\fusioncenter} }{ P^{[2]}_{\centeruser} \big| \tilde{h}_{\centeruser\fusioncenter} \big|^2 +\sigma^2_{\fusioncenter} }
\right),
\label{eq_R_ADMFI_C_xE}
\end{equation}
where $P^{[2]}_{\centeruser} \big| \tilde{h}_{\centeruser\fusioncenter} \big|^2$ denotes the residual interference (RI) power due to the imperfect cancellation of $x^{[2]}_{\centeruser}$ at $\fusioncenter$, and $\tilde{h}_{\centeruser\fusioncenter} \sim \mathcal{CN}(0,\xi_{\fusioncenter} {\pl}_{\centeruser\fusioncenter} (_{\centeruser\fusioncenter}) )$ with $\xi_{\fusioncenter} \!\in\! [0,1]$ being the RI level due to imperfect SIC at $\fusioncenter$ \cite{KaderLCOMM2017}.

In the event $\{ {\event}^{[\fusioncenter]}_{\uav\to\centeruser} \!:=\! 
     g^2_{\centeruser\fusioncenter} {\pl}_{\centeruser\fusioncenter}
        \!<\! g^2_{\uav\fusioncenter} {\pl}_{\uav\fusioncenter} \}$, which is~the complementary event of ${\event}^{[\fusioncenter]}_{\centeruser\to\uav}$, $\fusioncenter$ primarily decodes $\hat{x}_{\edgeuser}$ while treating $x^{[2]}_{\centeruser}$ as interference.
    In this case, $\fusioncenter$ decodes $\hat{x}_{\edgeuser}$ with the achievable rate given by
\begin{equation}
R^{[\fusioncenter]}_{\uav\to\centeruser}
=   \frac{1}{2}
\log_2\bigg(
	1 + \frac{ P_{\uav} g_{\uav\fusioncenter}^2 {\pl}_{\uav\fusioncenter} }{ P^{[2]}_{\centeruser} g_{\centeruser\fusioncenter}^2 {\pl}_{\centeruser\fusioncenter} + \sigma^2_{\fusioncenter} }
\bigg).
\label{eq_R_ADMFII_E_xE}
\end{equation}

Once $\hat{x}_{\edgeuser}$ is correctly decoded, $\fusioncenter$ then performs SIC in an attempt to remove $\hat{x}_{\edgeuser}$ from $y^{[2]}_{\fusioncenter}$, then decodes $x^{[2]}_{\centeruser}$. 
    Assuming imperfect SIC, $\fusioncenter$ decodes $x^{[2]}_{\centeruser}$ with the achievable rate given~by
\begin{equation}
R^{[\fusioncenter]}_{\centeruser}
=   \frac{1}{2}
\log_2\left(
	1 + \frac{ P^{[2]}_{\centeruser} g_{\centeruser\fusioncenter}^2 {\pl}_{\centeruser\fusioncenter} }{ P_{\uav} \big| \tilde{h}_{\uav\fusioncenter} \big|^2 + \sigma^2_{\fusioncenter} }
\right),
\label{eq_R_ADMFII_E_xC}
\end{equation}
where $P_{\uav} \big| \tilde{h}_{{\uav} {\fusioncenter}} \big|^2$ denotes the residual power due to the imperfect cancellation of $\hat{x}_{\edgeuser}$, and ${\tilde{h}_{\uav\fusioncenter} \sim \mathcal{CN}(0,\xi_{\fusioncenter} {\pl}_{\centeruser\fusioncenter} )}$.

    While $\uav$ is only required to correctly decode $x_{\edgeuser}$, $\fusioncenter$ is required to correctly decode both of the information signals of $\centeruser$ and $\edgeuser$.
Not only so, $\fusioncenter$ also needs to maintain fairness in the performance between $\hat{x}_{\edgeuser}$ and $x^{[2]}_{\centeruser}$.
    In ${\event}^{[\fusioncenter]}_{\centeruser\to\uav}$, $x^{[2]}_{\centeruser}$ is transmitted over a dominant uplink, and thus should be decoded first to avoid interference while decoding $\hat{x}_{\edgeuser}$.
In ${\event}^{[\fusioncenter]}_{\uav\to\centeruser}$, $\fusioncenter$ primarily decodes $\hat{x}_{\edgeuser}$ to meet $\edgeuser$'s quality-of-service (QoS) which in turn avoids extreme interference while $\fusioncenter$ decodes $x^{[2]}_{\centeruser}$.

In addition, when $\uav$ remains silent during the second phase, $\fusioncenter$ only receives $x^{[2]}_{\centeruser}$ from $\centeruser$.
As a result, the instantaneous rate at $\fusioncenter$ is given by
\begin{equation}
R^{[2]}_{\centeruser} = 
\frac{1}{2}
\log_2\bigg(
	1 + \frac{ P^{[2]}_{\centeruser} g_{\centeruser\fusioncenter}^2 {\pl}_{\centeruser\fusioncenter} }{ \sigma^2_{\fusioncenter} }
\bigg).
\label{eq_FC_xC2}
\end{equation}

{{\renewcommand{\arraystretch}{1.5}
\begin{table}[!t]
\centering
\caption{{SUCCESSFUL DECODING EVENTS at the FC.\label{tab:deci_FC}}}
\begin{tabular}{|c|c|c|c|}
\hline
{UAV's Decision} & {SINR Events} & {$x_\centeruser^{[2]}$} & {$\hat{x}_\edgeuser$}
\\\hline
\multirow{6}{*}{forward} & $\{ \event_{\centeruser\to\uav}^{[\fusioncenter]}, \bar{\success}_{\centeruser\to\uav}^{[\fusioncenter]} \}$ & Outage & Outage 
\\\cline{2-4}
& $\{ \event_{\centeruser\to\uav}^{[\fusioncenter]}, {\success}_{\centeruser\to\uav}^{[\fusioncenter]}, \bar{\success}_{\uav}^{[\fusioncenter]} \}$ & Outage & Non-Outage 
\\\cline{2-4}
& $\{ \event_{\centeruser\to\uav}^{[\fusioncenter]}, {\success}_{\centeruser\to\uav}^{[\fusioncenter]}, {\success}_{\uav}^{[\fusioncenter]} \}$ & Non-Outage & Non-Outage 
\\\cline{2-4}
 & $\{ \event_{\uav\to\centeruser}^{[\fusioncenter]}, \bar{\success}_{\uav\to\centeruser}^{[\fusioncenter]} \}$ & Outage & Outage 
\\\cline{2-4}
& $\{ \event_{\uav\to\centeruser}^{[\fusioncenter]}, {\success}_{\uav\to\centeruser}^{[\fusioncenter]},
\bar{\success}_{\centeruser}^{[\fusioncenter]} \}$ & Outage & Non-Outage 
\\\cline{2-4}
& $\{ \event_{\uav\to\centeruser}^{[\fusioncenter]}, {\success}_{\uav\to\centeruser}^{[\fusioncenter]},
{\success}_{\centeruser}^{[\fusioncenter]} \}$ & Non-Outage & Non-Outage 
\\\hline
\multirow{2}{*}{silent} & $\{ \bar{\success}_{\centeruser}^{[2]} \}$ & Outage & Outage
\\\cline{2-4}
& $\{ {\success}_{\centeruser}^{[2]} \}$ & Non-Outage & Outage
\\\hline
\end{tabular}\vspace{-0.4cm}
\end{table}}}

The FC's decoding events and the corresponding received SINR events are summarized in Table \ref{tab:deci_FC}, where the decoding event at $\fusioncenter$ is expressed as
    ${{\success}_{\centeruser}^{[2]} \!:=\! \{ R_{\centeruser}^{[2]} \!>\! R_{\threshold,\centeruser} \}}$, 
and the successful~decoding events at $\uav$ are
    ${{\success}_{\centeruser\to\uav}^{[\fusioncenter]}
    \!:=\! \{ R_{\centeruser\to\uav}^{[\fusioncenter]} \!>\! R_{\threshold,\centeruser} \}}$,
    ${{\success}_{\uav}^{[\fusioncenter]}
    \!:=\! \{ R_{\uav}^{[\fusioncenter]} \!>\! R_{\threshold,\edgeuser} \}}$,
    ${{\success}_{\uav\to\centeruser}^{[\fusioncenter]}
    \!:=\! \{ R_{\uav\to\centeruser}^{[\fusioncenter]} \!>\! R_{\threshold,\edgeuser} \}}$,
    and 
    ${{\success}_{\centeruser}^{[\fusioncenter]}
    := \{ R_{\centeruser}^{[\fusioncenter]} > R_{\threshold,\centeruser} \}}$.
The proposed ADM-aided CDRT-NOMA is summarized as Algorithm \ref{alg_ADM}. 
    It is noted that different transmission phases have different maximum power budgets for the following reasons. First, the second phase may involve transmission of only $\centeruser$, as opposed to the first phase, which always involves $\centeruser$ and $\edgeuser$. 
Second, $\uav$ and $\edgeuser$ have different architectures, where $\edgeuser$ is a ground UE while $\uav$ relays information from $\edgeuser$ to $\fusioncenter$. In addition, a generalized transmission model can be obtained by assuming different maximum power budgets.

Furthermore, when $\uav$ correctly decodes both $x_\edgeuser$ and $x_\centeruser^{[1]}$, $\uav$ just forwards $\hat{x}_\edgeuser$ to $\fusioncenter$. 
    This prevents $\fusioncenter$ from receiving three different types of information signals: $x_\centeruser^{[2]}$ and the mixture of $\hat{x}_\edgeuser$ and the regenerated version of $x_\centeruser^{[1]}$, denoted as $\hat{x}_{\centeruser}^{[1]}$. 
When $\fusioncenter$ receives more than two signals simultaneously, it is required to determine a more complex decoding architecture, which is beyond the scope of this paper.
    On the other hand, the decoding of $x_\centeruser^{[2]}$ and $\hat{x}_{\edgeuser}$ may suffer from additional interference from $\hat{x}_{\centeruser}^{[1]}$ and the RI power, which will eventually reduce the performance of the proposed system.

\vspace{-0.1cm}
\begin{algorithm} 
\caption{Proposed ADM for UAV-aided DUL CDRT-NOMA.} 
\label{alg_ADM}
\textbf{initialize}: 
Received signals at $\uav$, $y_{\uav}$, and $\fusioncenter$, $y_\fusioncenter$, in the first phase, ${\tt isUAVSilent} \!\leftarrow\! \textbf{false}$,
${\tt signalOut} \!\leftarrow\! \textbf{empty}$;

{\bf if }{${\success}^{[1]}_{\centeruser}$,}{ 
    \textbf{add} $x^{[1]}_{\centeruser}$ into ${\tt signalOut}$;
} \\
{\bf else}, {$\fusioncenter$ fails to decode $x^{[1]}_{\centeruser}$;} {\bf end if};

{\bf if }{${\event}^{[\uav]}_{\centeruser\to\edgeuser}$},
{\bf if}
    {${\success}^{[\uav]}_{\centeruser\to\edgeuser}$,}
        $\uav$ cancels $x^{[1]}_{\centeruser}$ with imperfect SIC; \\
\qquad {\bf if }{${\success}^{[\uav]}_{\edgeuser}$},
    {${\tt isUAVSilent} \leftarrow \textbf{false}$;} \\
\qquad {\bf else},
    {${\tt isUAVSilent} \leftarrow \textbf{true}$;} {\bf end if};\\
\quad {\bf else},
    {${\tt isUAVSilent} \leftarrow \textbf{true}$;} {\bf end if}; \\
{\bf elsif} {${\event}^{[\uav]}_{\edgeuser\to\centeruser}$}, 
{\bf if} {${\success}^{[\uav]}_{\edgeuser\to\centeruser}$,} 
    {${\tt isUAVSilent} \leftarrow \textbf{false}$;} \\
\quad {\bf else}, {${\tt isUAVSilent} \leftarrow \textbf{true}$;} {\bf end if}; {\bf end if};

{\bf if~not} {${\tt isUAVSilent}$} \\
\quad {\bf if} {${\event}^{[\fusioncenter]}_{\centeruser\to\uav}$},
{\bf if} {${\success}^{[\fusioncenter]}_{\centeruser\to\uav}$},
    {\textbf{add} $x^{[2]}_{\centeruser}$ into ${\tt signalOut}$;} {$\fusioncenter$ cancels $x^{[2]}_{\centeruser}$ with imperfect SIC;}\\
\qquad\quad {\bf if} {${\success}^{[\fusioncenter]}_{\uav}$,}
    {\textbf{add} $\hat{x}_{\edgeuser}$ into ${\tt signalOut}$;} \\
\qquad\quad {\bf else}, {$\fusioncenter$ fails to decode $\hat{x}_{\edgeuser}$;} {\bf end if}; \\
\qquad {\bf else}, {$\fusioncenter$ fails to decode $x^{[2]}_{\centeruser}$ and $\hat{x}_{\edgeuser}$;} {\bf end if}; \\
\quad {\bf elsif} {${\event}^{[\fusioncenter]}_{\uav\to\centeruser}$},
    {\bf if} {${\success}^{[\fusioncenter]}_{\uav\to\centeruser}$},
    {\textbf{add} $\hat{x}_{\edgeuser}$ into ${\tt signalOut}$;} {$\fusioncenter$ cancels $\hat{x}_{\edgeuser}$ with imperfect SIC;} \\
\qquad\quad {\bf if} {${\success}^{[\fusioncenter]}_{\centeruser}$},
    {\textbf{add} $x^{[2]}_{\centeruser}$ into ${\tt signalOut}$;}\\
\qquad\quad {\bf else}, {$\fusioncenter$ fails to decode $\hat{x}_{\edgeuser}$;} {\bf end if}; \\
\qquad {\bf else}, {$\fusioncenter$ fails to decode $x^{[2]}_{\centeruser}$ and $\hat{x}_{\edgeuser}$;} {\bf end if}; \\
{\bf else}, {\bf if} {${\success}^{[2]}_{\centeruser}$},
    {\textbf{add} $x^{[2]}_{\centeruser}$ into ${\tt signalOut}$;} \\
\quad {\bf else}, {$\fusioncenter$ fails to decode $x^{[2]}_{\centeruser}$ and $\hat{x}_{\edgeuser}$}; {\bf end if}; {\bf end if}; \\
\textbf{output}: ${\tt signalOut}$.
\end{algorithm}	

\section{Performance Analysis for Hovering UAV}
\subsection{G2G and A2G/G2A Channels Characterization}
In this subsection, we determine the statistical property of the G2G and A2G/G2A channel power gains. First, we denote
$
{\psi_{\centeruser\fusioncenter}
    \!=\!   g_{\centeruser\fusioncenter}^2 {\pl}(d_{\centeruser\fusioncenter})}$ and
$
{\phi_{\sf XY}
    \!=\!   g_{\sf XY}^2 {\pl}_{\sf XY}}$ as the power gain of the G2G and the A2G/G2A links, respectively. Using the property $f_{aX}(x)=\frac{f_{X}(x/a)}{a}$ for $a > 0$ and $X$ is a RV, the PDF of $\psi_{\centeruser\fusioncenter}$ is obtained as
\begin{align}
{f}_{ \psi_{\centeruser\fusioncenter} }(x)
    &=  \alpha 
        \sum_{l=1}^{m_{\centeruser\fusioncenter}-1}
        \zeta(l) \frac{ x^{l} }{ l! } e^{-\frac{\beta-\delta}{{\pl}(d_{\centeruser\fusioncenter})} x}
    +   \alpha 
        e^{ -\frac{\beta-\delta}{{\pl}(d_{\centeruser\fusioncenter})} x }.
\label{eq_apx_pdf_hCF}
\end{align}
In addition, the PDF of $\phi_{\sf XY}$ is obtained as
\begin{align}
{f}_{ \phi_{\sf XY} }(x)
    &= p_{\sf XY}^\los
    \frac{ \left(
       \frac{ m_{\sf XY} }{ \pl^\los_{\sf XY} }
    \right)^{-m_{\sf XY}} }{ (m_{\sf XY}-1)! } 
    x^{m_{\sf XY}-1}
    e^{ -\frac{ m_{\sf XY} }{ \pl^\los_{\sf XY} } x }
    \nonumber\\ 
&\quad+  \frac{ p_{\sf XY}^\nlos }{  \pl^\nlos_{\sf XY}  } 
    e^{ -\frac{x}{\pl^\nlos_{\sf XY}} },
x > 0.
\label{eq_apx_pdf_hXY}
\end{align}

Let us denote $\varphi_{\sf XY} \triangleq {\big|\tilde{h}_{\sf XY}\big|^2} $ as the fraction of RI power due to imperfect SIC; thus, the PDF of $\varphi_{\sf XY}$ is obtained as
\begin{align}
f_{ \varphi_{\sf XY} }(x)
    = \frac{ 1 }{ \xi_{\sf Y} \overline{\pl}_{\sf XY} }
    e^{ -\frac{ x }{ \xi_{\sf Y}\overline{\pl}_{\sf XY} } },
    ~x > 0,
\end{align}
where 
$
\overline{\pl}_{\sf XY}
    \triangleq   p_{\sf XY}^\los {\pl}^{\los}_{\sf XY} 
    \!+\!   p_{\sf XY}^\nlos {\pl}^{\nlos}_{\sf XY}.
$
By combining \eqref{eq_apx_pdf_hCF} and \eqref{eq_apx_pdf_hXY} into a unified expression, we then obtain the following Lemma. 
\begin{Lemma}
\label{lem_1}
The PDF and the CDF of a $(K \!+\! 1)$ Mixture of Gamma (MG) distributions, each with the mixing proportion $\chi(k)$, shape $\mu(k)$ and scale $\Omega(k)$, are expressed~as \cite{AtapattuTWC2011}
\begin{align}
{f}_{\gamma}(\Theta; x) 
    &\!=\!   \sum_{k=1}^{K+1}
		\chi(k)
		\frac{ \Omega(k)^{-\mu(k)} }{ (\mu(k)-1)! }			 
	 	x^{ \mu(k)-1 } e^{ -\frac{x}{\Omega(k)} },
\label{eq_pdf_G_final} \\
{F}_{\gamma}(\Theta; x)
	&\!=\!	1 -\!\! \sum_{k=1}^{ {K+1} }
	\frac{\chi(k) }{(\mu(k) \!-\! 1)!}
	\Gamma\left(
		\mu(k), \frac{x}{\Omega(k)}
	\right),
\label{eq_cdf_G_final} 
\end{align}
respectively, 
where $\Gamma(a,x)$ is the upper incomplete Gamma function \cite{Gradshteyn2007}. Herein, the argument $\Theta \triangleq (\theta_1, \dots, \theta_{K+1})$ contains parameters of all $(K+1)$ mixtures, in which
${ \theta_k = (\chi(k), \Omega(k), \mu(k)) }$ is the parameters of the $k$-th~mixture. 
\end{Lemma}
By adopting Lemma \ref{lem_1}, we can represent the G2G, A2G/G2A channel power gains and the portion of RI power as MG variates.
    It is worth noting that the results in the following sections can be adopted to characterize various systems over fading channels having MG-like distribution, e.g., $\kappa-\mu$ shadowed, Rician shadowed, generalized-$K$, and Rician \cite{EspinosaGLOBECOM2019}.
Hereafter, we denote ${X \sim {\rm MG}(\Theta)}$ as the RV $X$ has MG distribution with parameters $\Theta$. 
    The distributions of $\psi_{\centeruser\fusioncenter}$, $\phi_{\sf XY}$, and $\varphi_{\sf XY}$ can be represented by MG distributions with the parameters $\Theta_{\centeruser\fusioncenter}$ ($K = m_{\centeruser\fusioncenter}$), $\Theta_{\sf XY}$ ($K = 1$), and $\hat{\Theta}_{\sf XY}$  ($K = 0$) shown in Table \ref{tab:parameters}. 
    Note that the shape factors $m_{\sf XY}$ in Table \ref{tab:parameters} are assumed to have integer values. To deal with real-valued shapes, we introduce the following Lemma.   
{\renewcommand{\arraystretch}{1.5}
\begin{table}
    \centering
    \caption{{Parameters of MG distribution. \label{tab:parameters}}}
    \begin{tabular}{|c|c|c|}
    \hline
        {} & {$1 \le k \le K$} &{$k = K+1$} \\
    \hline\hline
        {$[\Theta_{\centeruser\fusioncenter}]_k$} &
        {$\Big( \frac{\alpha \zeta(k)}{(\beta-\delta)^{k+1}}, k+1, \frac{{\pl}_{\centeruser\fusioncenter}}{\beta-\delta} \Big)$}&
        {$\Big( \frac{\alpha}{\beta-\delta}, 1, \frac{{\pl}_{\centeruser\fusioncenter}}{\beta-\delta} \Big)$}
    \\\hline
        {$[\Theta_{\sf XY}]_k$} & $\Big( p_{\sf XY}^\los, m_{\sf XY}, {\pl}^{\los}_{\sf XY} \Big)$ & 
        $\Big( p_{\sf XY}^\nlos, 1, {\pl}^{\nlos}_{\sf XY} \Big)$ \\
    \hline
        {$[\hat{\Theta}_{\sf XY}]_k$} & $-$ & {$\Big( 1, 1, \xi_{\sf Y} \overline{\pl}_{\sf XY} \Big)$} \\
    \hline
    \end{tabular}
\vspace{-0.5cm}
\end{table}}

\begin{Lemma} \label{lemma_realValued_Gamma}
Given a Gamma distributed RV ${g \!\sim\! {\Gamma}(m,\theta)}$ with real-valued shape $m$, $m \ge \frac{1}{2}$, and scale $\sigma = \frac{1}{m}$, the PDF of ${g}$ can be accurately formulated as
\begin{equation}
\label{eq:PDF_Gamma_Aprx_fin}
f_{g}(x) =  
    \sum_{t=1}^{\lfloor m \rfloor} \upzeta_{t,M}
        \frac{ m^t }{(t-1)!}
        x^{t-1} e^{-m x}
    + \sum_{t=1}^{M} \upzeta_{t}
        \frac{1}{\sigma_t} e^{-\frac{x}{\sigma_t}},
\end{equation}
where  ${\upzeta_t \!\triangleq\! 
	\frac{w_t}{ ( 1-\frac{1}{m\theta_t})^{\lfloor m \rfloor}}}$, 
and~${ \upzeta_{t,M} \!\triangleq\! -\sum_{k=1}^{M} 
	\frac{ 1 }{ m \sigma_k }
	\frac{ w_k }{( 1-\frac{1}{m \sigma_k} )^{\lfloor m \rfloor-t+1}} }$.
\end{Lemma}
\begin{IEEEproof}
See Appendix \ref{apx_realValued_Gamma}.
\end{IEEEproof}

Hence, for $m_{\sf XY}$ having real values, the parameters of $\phi_{\sf XY}$ in Table \ref{tab:parameters} are rewritten as
\begin{align}
[\Theta_{\sf XY}]_{t}
    =   \left\{
    \begin{array}{l}
        \left(
            p^\los_{\sf XY} \zeta_t, 1,
            \pl^\los_{\sf XY} \sigma_{t}
        \right), \qquad\quad {1 \le t \le M}, \vspace{5pt}\\
        \left(
            p^\los_{\sf XY} \zeta_{t-M,M}, t - M,
           { \pl^\los_{\sf XY} }/{m_{\sf XY}}
        \right), \\
            \qquad\qquad\quad~
            {M+1 \le t \le M+\lfloor m_{\sf XY} \rfloor}, \vspace{0.5pt}\\
        \left(
            p^\nlos_{\sf XY}, 1, \pl^\nlos_{\sf XY}
        \right),~ {t = M \!+\! \lfloor m_{\sf XY}\rfloor} \!+\! 1,  \\
    \end{array}
    \right.\!
\end{align}
where $K = M + \lfloor m_{\sf XY} \rfloor$.

\vspace{-5pt}
\subsection{Preliminary}

In this subsection, we introduce important integrals to assist with the forthcoming analysis. 
These integrals will significantly simplify the analysis. 
Specifically, given~that
${\vec{x}_n \triangleq (x_n,\dots,x_1)}$ is the variables of integration, 
${\vec{w}_n \triangleq (p_n x_{n+1} + q_{n},\dots,p_1 x_2 + q_1 )}$ is the lower limits of the multiple integrals, 
${\vec{p}_n \triangleq (p_n,\dots,p_1)}$,
${p_i \ge 1}$, ${\forall i \in [1,n]}$, and
${\vec{q}_n \!\triangleq\! (q_n,\dots,q_1)}$, $q_i \!\ge\! 0$, $\forall i \!\in\! [1,n]$, the integral is defined~as
\begin{align}
\mathpzc{W}_n(\Theta_n, \dots, \Theta_1, {\bf p}_n, {\bf q}_n) = 
    \int\limits_{ D_n }
    \left[ 
        \prod_{t=1}^{n}{ {f}_{{\gamma}_t}(x_t) }
    \right]
    {\rm d} \vec{x}_n,
\label{eq_In_def}
\end{align}
where
    ${ D_n \!\triangleq\! \{ {\bf x}_n : [{\bf x}_n]_j \ge [{\bf w}_n]_j, 1 \!\le\! j \!\le\! n \} }$ is the integration domain and 
    ${\gamma}_t \!\sim\! {\rm MG}(\Theta_t)$, for ${t \!\in\! [1,n]}$, are statistically independent MG variates with ${[\Theta_t]_k \!=\! (\chi_t(k), \mu_t(k), \Omega_{t}(k))}$.
It is noted that \eqref{eq_In_def} represents the integral of the joint multiple MG-based distribution, which is $\prod_{t = 1}^{n}{ {f}_{\gamma_t}(x_t) }$, over the integration domain $D$.
    For convenience, let the argument of $\mathpzc{W}_n$ in \eqref{eq_In_def} be ${\bf W}_{n} = (\Theta_n, \dots, \Theta_1,
    {\bf p}_n, {\bf q}_n)$,
the integral in \eqref{eq_In_def} can be derived from the following probability
\begin{align}
\mathpzc{W}_n({\bf W}_{n}) 
= \Pr\!\left[
    \bigcap_{t=1}^{n}
        {\gamma}_t > [{\bf p}_n]_{t} {\gamma}_{t} \!+\! [{\bf q}_n]_{t}
\right].
\label{eq_In_def_prob}
\end{align}
In general, $\mathpzc{W}_n(\cdot)$ can be derived in closed-form expression as
\begin{align}
\label{eq:Wn_CFE}
\mathpzc{W}_n({\bf W}_{n}) 
    =  \mathop{\widetilde{\sum}}\limits_{ n }
    \Xi_n
	(\Lambda_n)^{\kappa_n}
	\Gamma\bigg(
		\kappa_n, \frac{ w_{n} }{ \Lambda_n } 
	\bigg) 
	\Phi_n,
\end{align}
where $\mathop{\widetilde{\sum}}_{ n }$ represents the short-hand notations for the multiple summation
    $ \sum_{k_1=1}^{K_1+1} 
	\sum_{k_1=1}^{K_1+1}
	\cdots
	\sum_{k_n=1}^{ {K_n+1} }
	\allowbreak
    \sum_{i_1=0}^{\kappa_1-1}
	\allowbreak
	\sum_{ {j_1+l_1=i_1} }
	\allowbreak
	\cdots
    \sum_{i_{n-1}=0}^{\kappa_{n-1}-1}
    \sum_{ j_{n-1}+l_{n-1}=i_{n-1} }$, and
\begin{align}
&\qquad 
\Lambda_t
	\triangleq \bigg( \frac{1}{\Omega_t(k_t)} + \frac{p_{t-1}}{\Lambda_{t-1}} \bigg)^{-1},~
\kappa_t 
	\!\triangleq\! \mu_t(k_t) + l_{t-1},~\forall t \ge 2,
\nonumber\\
&\qquad
\Xi_t
	\triangleq \frac{ \chi_t(k_t) }{ (\mu_t(k_t)-1)! } 
	\Omega_t(k_t)^{-\mu_t(k_t)},~\forall t \ge 1,
\nonumber\\
&\Phi_n
    \triangleq
    \prod_{t=1}^{n-1} \Psi_{t},~
\Psi_{t} \triangleq
    \frac{ \Xi_t }{ (\Lambda_t)^{j_t} } 
    \frac{ (q_t)^{j_t} }{ j_t! }
    \frac{ (p_t)^{l_t} }{ l_t! }
    \frac{\Gamma(\kappa_t) e^{-\frac{q_t}{\Lambda_t}}}{(\Lambda_t)^{-\kappa_t}},
~\forall n \ge 2,
\nonumber
\end{align}
in which $\Lambda_1 = \Omega_1(k_1)$ and $\kappa_1 = \mu_1(k_1)$.
\begin{IEEEproof}
See Appendix \ref{apx_lemma_2}.
\end{IEEEproof}

In addition, by using similar steps in Appendix \ref{apx_lemma_2}, we obtain the following function
\begin{align}
\mathpzc{w}_n({\bf W}_{n}) 
&=
\Pr\!\Bigg[
    \bigcap_{t=1}^{n-1}
    {\gamma}_t > [{\bf p}_n]_{t} {\gamma}_{t} \!+\! [{\bf q}_n]_{t}, \nonumber\\
&\qquad\quad
    {\gamma}_n < [{\bf p}_n]_{n} {\gamma}_{n} \!+\! [{\bf q}_n]_{n}
\Bigg], \\
&=  \mathop{\widetilde{\sum}}\limits_{ n }
    \Xi_n
    (\Lambda_n)^{\kappa_n}
    \gamma\bigg(
        \kappa_n, \frac{ w_{n} }{ \Lambda_n } 
    \bigg) 
    \Phi_n.
\label{eq:wn_CFE}
\end{align}  

\noindent When $\exists v \!\in\! [1,n-1]$ that $[{\bf p}_n]_{v} \!=\! 0$, we have
\begin{align}
\mathpzc{W}_n({\bf W}_{n}) 
    &= \mathpzc{W}_{n-v-1}({\bf W}_{n,v+1}) 
    \mathpzc{W}_{v}({\bf W}_{v}), \\
\mathpzc{w}_n({\bf W}_{n}) 
    &= \mathpzc{w}_{n-v-1}({\bf W}_{n,v+1}) 
    \mathpzc{W}_{v}({\bf W}_{v}),
\end{align}
where ${\bf W}_{j,i} =  (\Theta_j, \dots, \Theta_i,
    [{\bf p}_n]_j,\dots,[{\bf p}_n]_i, [{\bf q}_n]_j,\dots,[{\bf q}_n]_i)$.

\noindent  When $\exists v \!\in\! [1,n-1]$ that $[{\bf q}_n]_v = 0$, $j_{v} = 0$ and $l_{v} = i_{v}$.

\noindent When $n = 1$, the self-defined integral is simplified as
\begin{align}
\mathpzc{W}_1({\bf W}_1) = F^c_{\gamma_1}({\bf w}_1).
\end{align}
Next, we consider delay-limited services at both $\centeruser$ and $\edgeuser$ and evaluate the outage probability to examine their performance.
    Let us define ${R}_{\sf th,C}$ as the target transmission rate of both $x^{[1]}_{\centeruser}$ and $x^{[2]}_{\centeruser}$, and ${R}_{\sf th,E}$ as that of both $x_{\edgeuser}$ and $\hat{x}_{\edgeuser}$. 
    If the instantaneous achievable rate falls below the corresponding target transmission rate, an outage event is said to occur.
The end-to-end (e2e) outage probability (OP) of each UE is examined in the following subsections.~
\vspace{-5pt}
\subsection{e2e OP of $\edgeuser$}~
In this subsection, we examine the e2e OP of $\edgeuser$, denoted as ${\rm OP}_{\edgeuser,{\rm e2e}}$.
    Since the transmission of $x_{\edgeuser}$ from $\edgeuser$ to $\fusioncenter$ 
    is assisted by the decode-and-forward (DF) UAV, 
    ${\rm OP}_{\edgeuser,{\rm e2e}}$ relies on the probability that $\uav$ accurately decodes $x_{\edgeuser}$.
According to Table \ref{tab:deci_FC}, the event that $\uav$ correctly decodes $x_{\edgeuser}$ is given by
\begin{align}
{\correct}^{[\uav]}_{\edgeuser} := 
    \{ {{\event}^{[\uav]}_{\centeruser\to\edgeuser}}, {{\success}^{[\uav]}_{\centeruser\to\edgeuser}}, {{\success}^{[\uav]}_{\edgeuser}} \}
    \cup 
    \{ {{\event}^{[\uav]}_{\edgeuser\to\centeruser}}, {{\success}^{[\uav]}_{\edgeuser\to\centeruser}} \}.
\end{align}

Let us denote ${\correct}^{[\uav]}_{\centeruser \to \edgeuser} \!:=\! \{ {\event}^{[\uav]}_{\centeruser\to\edgeuser}, {\success}^{[\uav]}_{\centeruser\to\edgeuser}, {\success}^{[\uav]}_{\edgeuser} \}$
    and ${\correct}^{[\uav]}_{\edgeuser \to \centeruser} \!:=\! \{ {{\event}^{[\uav]}_{\edgeuser\to\centeruser}}, {{\success}^{[\uav]}_{\edgeuser\to\centeruser}} \}$, the probability of non-outage at $\uav$ can be formulated as
\begin{align}
P^{[\uav]}_{\edgeuser}
    =  \Pr\left[ {\correct}^{[\uav]}_{\edgeuser} \right]
    =  \Pr\left[ {\correct}^{[\uav]}_{\centeruser \to \edgeuser} \right]
    + \Pr\left[ {\correct}^{[\uav]}_{\edgeuser \to \centeruser} \right].
\label{eq_Pout_uav_sim}
\end{align}

The closed-form expressions of 
    $P^{[\uav]}_{\centeruser \to \edgeuser} \!\triangleq\! \Pr[ {\correct}^{[\uav]}_{\centeruser \to \edgeuser} ]$ and 
    $P^{[\uav]}_{\edgeuser \to \centeruser} \!\triangleq\! \Pr[ {\correct}^{[\uav]}_{\edgeuser \to \centeruser} ]$ are provided by the following Lemmas.

\begin{Lemma} \label{lem_closedForm_P1}
The closed-form expression of $P^{[\uav]}_{\centeruser \to \edgeuser}$ is obtained~as
\begin{align}
P^{[\uav]}_{\centeruser \to \edgeuser} 
    \!=\! 
    \left\{
    \begin{array}{l}
        \mathpzc{W}_3(\hat{\Theta}_3, \Theta_2, \Theta_1, 0, \alpha_2, \alpha_1, {\cal A}, a_2, a_1) \\
        \quad- \mathpzc{w}_1(\hat{\Theta}_3, 0, {\cal A})
        \mathpzc{W}_2({\Theta}_2, \Theta_1, 0, \alpha_1, A_1, a_1) \\
        + \mathpzc{W}_3(\hat{\Theta}_3, \Theta_2, \Theta_1, 0, \alpha_2, 1, {\cal A}, a_2, 0) \\
        \quad- \mathpzc{w}_1(\hat{\Theta}_3, 0, {\cal A})
        \mathpzc{W}_2({\Theta}_2, \Theta_1, 1, A_1, 0), \\
        \hfill { \text{ if } \alpha_1 < 1 \text{ and } A_1 > a_2}, \\
        \mathpzc{W}_3(\hat{\Theta}_3, \Theta_2, \Theta_1, 0, \alpha_2, 1,
        0, a_2, 0) \\
        \quad+ \mathpzc{w}_1(\hat{\Theta}_3, 0, {\cal A})
        \mathpzc{W}_2({\Theta}_2, \Theta_1, 0, \alpha_1, A_1, a_1), \\
        \hfill { \text{ if } \alpha_1 \ge 1}.
    \end{array}
    \right.
    \label{eq:Lem3_final}
\end{align}
where $\Theta_3 \!\triangleq\! \hat{\Theta}_{\centeruser\uav}$,
      $\Theta_2 \!\triangleq\! {\Theta}_{\edgeuser\uav}$,
      $\Theta_1 \!\triangleq\! {\Theta}_{\centeruser\uav}$,
    ${a_1 \!\triangleq\! \frac{ \tau_{\sf th, C} }{ \bar{\gamma}_{\centeruser} }}$,
    ${a_2 \!\triangleq\! \frac{ \tau_{\sf th, E} }{ \bar{\gamma}_{\edgeuser} }}$,
    ${\alpha_1 \!\triangleq\! \frac{ \bar{\gamma}_{\edgeuser} \tau_{\sf th, C} }{ \bar{\gamma}_{\centeruser} }}$, 
    ${\alpha_2 \!\triangleq\! \frac{ \bar{\gamma}_{\centeruser} \tau_{\sf th, E} }{ \bar{\gamma}_{\edgeuser} }}$,
    ${A_1 \!\triangleq\! \frac{a_1}{1-\alpha_1}}$,
    ${A_2 \!\triangleq\! \frac{a_2}{1-\alpha_2}}$, and
    ${\mathcal{A} \!\triangleq\! \frac{A_1 - a_2}{\alpha_2}}$.
\end{Lemma}
\begin{IEEEproof}
	See Appendix \ref{apx:B}.
\end{IEEEproof}

\begin{Lemma} \label{lem:2}
The closed-form expression of $P^{[\uav]}_{\edgeuser \to \centeruser}$ is obtained~as
\begin{align}
P^{[\uav]}_{\edgeuser \to \centeruser}
    =   \left\{
    \begin{array}{lc}
        \mathpzc{W}_2(\Theta_2, \Theta_1, \alpha_2, a_2, 0, A_1) \\
        \quad- \mathpzc{W}_2(\Theta_2, \Theta_1, \alpha_2, a_2, 0, 0) \\
        \quad+ \mathpzc{W}_2( \Theta_2, \Theta_1, 1, 0, 0, A_1), &
        { \text{ if } \alpha_2 < 1},\\
        \mathpzc{W}_1(\Theta_2, \Theta_1, \alpha_2, a_2, 0, 0), &      
        { \text{ if } \alpha_2 \ge 1}.
    \end{array}
    \right.
\label{eq:Lem4_final}
\end{align}
where $\Theta_2 \triangleq \Theta_{\centeruser\uav}$ and 
    $\Theta_1 \triangleq \Theta_{\edgeuser\uav}$.
\end{Lemma}

\begin{IEEEproof}
	See Appendix \ref{apx:C}.
\end{IEEEproof}

The proposed system adopts AMDs to decode the transmitted signal of $\edgeuser$, where the event of correctly decoding $\edgeuser$'s information signal is formulated as
\begin{align}
{\correct_\edgeuser^{[\fusioncenter]}}
    := \{ 
        {\event}^{[\fusioncenter]}_{\centeruser\to\uav}, {\success}^{[\fusioncenter]}_{\centeruser\to\uav}, {\success}^{[\fusioncenter]}_{\uav}
    \}
        \cup 
    \{
        {\event}^{[\fusioncenter]}_{\uav\to\centeruser}, {\success}^{[\fusioncenter]}_{\uav\to\centeruser}
    \}.
\label{eq_event_E}
\end{align}

By denoting ${{\correct}^{[\fusioncenter]}_{\edgeuser,\centeruser \to \uav} \!:=\! \{ {\event}^{[\fusioncenter]}_{\centeruser \to \uav}, {\success}^{[\fusioncenter]}_{\centeruser \to \uav}, {\success}^{[\fusioncenter]}_{\uav} \}}$
    and ${{\correct}^{[\fusioncenter]}_{\edgeuser,\uav \to \centeruser} \!:=\! \{ {{\event}^{[\fusioncenter]}_{\uav\to\centeruser}}, {{\success}^{[\fusioncenter]}_{\uav\to\centeruser}} \}}$; the  probability of correct decoding of $\hat{x}_\edgeuser$ at $\fusioncenter$ is formulated as
\begin{align}
P^{[\fusioncenter]}_{\edgeuser}
    =  \Pr\left[ {\correct}^{[\fusioncenter]}_{\edgeuser} \right]
    =  \Pr\left[ {\correct}^{[\fusioncenter]}_{\edgeuser,\centeruser \to \uav} \right]
        + \Pr\left[ {\correct}^{[\fusioncenter]}_{\edgeuser,\uav \to \centeruser} \right].
\label{eq_Pout_FC_sim}
\end{align}

The  probability $P^{[\fusioncenter]}_{\edgeuser}$ also comprises of two probabilities, 
    $P^{[\fusioncenter]}_{\edgeuser,\centeruser \to \uav} \!\triangleq\! \Pr[ {\correct}^{[\fusioncenter]}_{\edgeuser,\centeruser \to \uav} ]$ and 
    $P^{[\fusioncenter]}_{\edgeuser,\uav \to \centeruser} \!\triangleq\! \Pr[ {\correct}^{[\fusioncenter]}_{\edgeuser,\uav \to \centeruser} ]$, which corresponds to two possible~decoding orders determined by ${\event}^{[\fusioncenter]}_{\centeruser\to\uav}$ and ${\event}^{[\fusioncenter]}_{\uav\to\centeruser}$, respectively.
The closed-form expressions of $P^{[\uav]}_{\edgeuser,\centeruser \to \uav}$ and $P^{[\uav]}_{\edgeuser,\uav \to \centeruser}$ are provided by the following Lemma.
\begin{Lemma}
The closed-form expressions of $P^{[\fusioncenter]}_{\edgeuser,\centeruser \to \uav}$ and $P^{[\fusioncenter]}_{\edgeuser,\uav \to \centeruser}$ are analogous to $P^{[\uav]}_{\centeruser \to \edgeuser}$ and $P^{[\uav]}_{\edgeuser \to \centeruser}$, respectively. 
    Specifically, given that ${b_1 \triangleq \frac{ 1 }{ \bar{\gamma}_{\centeruser} } \tau_{\sf th, C}}$,
    ${b_2 \triangleq \frac{ 1 }{ \bar{\gamma}_{\uav} } \tau_{\sf th, E}}$,
    ${\beta_1 \triangleq \bar{\gamma}_{\uav} b_1}$,
    ${\beta_2 \triangleq  \bar{\gamma}_{\centeruser} b_2}$,
    ${B_1 \triangleq \frac{b_1}{1-\beta_1}}$,
    ${B_2 \triangleq \frac{b_2}{1-\beta_2}}$, and
    ${\mathcal{B} \triangleq \frac{B_1 - b_2}{\beta_2}}$, we obtain $P^{[\fusioncenter]}_{\centeruser \to \edgeuser} $ and
    $P^{[\fusioncenter]}_{\edgeuser \to \centeruser} $ by applying the following substitutions in the proof of Lemma \ref{lem_closedForm_P1} and Lemma~\ref{lem:2},
\begin{align}
\begin{array}{cccc}
\Theta_1 := \Theta_{\centeruser\fusioncenter}, &
\Theta_1 := \Theta_{\fusioncenter\uav}, &
\hat{\Theta}_3 := \hat{\Theta}_{\centeruser\fusioncenter}, \\
a_1 := b_1, & A_1 := B_1, & \alpha_1 := \beta_1, & {} \\
a_2 := b_2, & A_2 := B_2, & \alpha_2 := \beta_2, & \mathcal{A} := \mathcal{B}.
\end{array}\!\!\!
\label{eq_arg_P1P2_to_P3P4}
\end{align}
\end{Lemma}
According to Table \ref{tab:deci_FC}, the e2e OP of $\edgeuser$ is formulated as
\begin{align}
{\rm OP}_{\edgeuser,{\rm e2e}}
=   \Pr\left[ \bar{\correct}^{[\uav]}_{\edgeuser} \right]  
    + \Pr\left[ {\correct}^{[\uav]}_{\edgeuser} \right] 
    \Pr\left[ \bar{\correct}^{[\fusioncenter]}_{\edgeuser} \mid {\correct}^{[\uav]}_{\edgeuser} \right].
\label{eq_OPe2e_F_sim}
\end{align}

\begin{Theorem} \label{theo_OP_xE}
The e2e OP of $\edgeuser$ can be derived as
\begin{align}
{\rm OP}_{\edgeuser,{\rm e2e}}
    \!=\!  1 \!-\! \left( P^{[\uav]}_{\centeruser \to \edgeuser} 
        + P^{[\uav]}_{\edgeuser \to \centeruser} \right) 
    \left( P^{[\fusioncenter]}_{\edgeuser,\centeruser \to \edgeuser} 
        + P^{[\fusioncenter]}_{\edgeuser,\edgeuser \to \centeruser} \right).
\label{eq_OP_F_xE}
\end{align}
\end{Theorem}

\begin{IEEEproof}
Following Table \ref{tab:deci_FC} and Table \ref{tab:deci_UAV}, the events ${\correct}^{[\uav]}_{\edgeuser}$ and ${\correct}^{[\fusioncenter]}_{\edgeuser}$ are independent, thus \eqref{eq_OPe2e_F_sim} can be rewritten as
\begin{align}
{\rm OP}_{\edgeuser,{\rm e2e}}
    =   1  - \Pr\left[{\correct}^{[\uav]}_{\edgeuser} \right] 
    \Pr\left[ {\correct}^{[\fusioncenter]}_{\edgeuser} \right],
\end{align}
where $\Pr[{\correct}^{[\uav]}_{\edgeuser}]$ and $\Pr[{\correct}^{[\fusioncenter]}_{\edgeuser}]$ are given by \eqref{eq_Pout_uav_sim} and \eqref{eq_Pout_FC_sim} respectively. 
    This completes the proof of Theorem \ref{theo_OP_xE}.
\end{IEEEproof}
\vspace{-5pt}
\subsection{e2e OPs of $\centeruser$}

Recalling that $\centeruser$ transmits different information during the first and second phases. 
    According to Table \ref{tab:deci_FC}, the event that $\fusioncenter$ correctly decodes $x^{[2]}_{\centeruser}$ is formulated as
\begin{align}
    \{ 
    {\correct}^{[\uav]}_\edgeuser,
    \{
        \{ {\event}^{[\fusioncenter]}_{\centeruser\to\uav}, 
            {\success}^{[\fusioncenter]}_{\centeruser\to\uav} \} \cup 
        \{ {\event}^{[\fusioncenter]}_{\uav\to\centeruser}, 
            {\success}^{[\fusioncenter]}_{\uav\to\centeruser}, 
            {\success}^{[\fusioncenter]}_{\centeruser} \}
    \}
    \}
    \cup \{ \bar{\correct}^{[\uav]}_\edgeuser,
            {\success}^{[2]}_{\centeruser} \}.
\end{align}

By denoting $\correct^{[\fusioncenter]}_{\centeruser,\uav \to \centeruser} 
        \!:=\! \{ {\event}^{[\fusioncenter]}_{\uav\to\centeruser}, 
        {\success}^{[\fusioncenter]}_{\uav\to\centeruser}, 
        {\success}^{[\fusioncenter]}_{\centeruser} \} $,
    $\correct^{[\fusioncenter]}_{\centeruser,\centeruser \to \uav} \!:=\! \{ {\event}^{[\fusioncenter]}_{\centeruser\to\uav}, 
        {\success}^{[\fusioncenter]}_{\centeruser\to\uav} \}$, and
$\correct^{[\fusioncenter]}_{\centeruser_2} 
    := \{ \correct^{[\fusioncenter]}_{\centeruser,\centeruser \to \uav} \cup 
    \correct^{[\fusioncenter]}_{\centeruser,\centeruser \to \uav} \} $, 
the e2e OP of $\centeruser$ in the second phase can be formulated as
\begin{align}
{\rm OP}^{[2]}_{\centeruser,{\rm e2e}}
    =  \Pr\left[ {\correct}^{[\uav]}_\edgeuser \right] 
        \Pr\left[ \correct^{[\fusioncenter]}_{\centeruser_2} \right] 
    + \Pr\left[ \bar{\correct}^{[\uav]}_\edgeuser \right] 
        \Pr\left[ {\success}^{[2]}_{\centeruser} \right].
\label{eq_Pout_fc_xC_sim}
\end{align}

\begin{Theorem} \label{theo:4}
The closed-form expression of ${\rm OP}^{[2]}_{\centeruser,{\rm e2e}}$ is obtained~ as
\begin{align}
{\rm OP}^{[2]}_{\centeruser,{\rm e2e}}
    &=   \left( P^{[\uav]}_{\centeruser \to \edgeuser} 
        + P^{[\uav]}_{\edgeuser \to \centeruser} \right)
        \left(1 - \left( P^{[\fusioncenter]}_{\centeruser,\uav \to \centeruser}   
            + P^{[\fusioncenter]}_{\centeruser,\centeruser \to \uav} \right) \right) \nonumber\\
    &\quad+   
        \left(1 - \left(P^{[\uav]}_{\centeruser \to \edgeuser} 
            + P^{[\uav]}_{\edgeuser \to \centeruser} \right) \right) P^{[2]}_{\centeruser},
\label{eq_Pout_fc_xC_ana}
\end{align}
where $P^{[2]}_{\centeruser}$ is given by \eqref{eq_varrho_7_int}. 
\end{Theorem}

\begin{IEEEproof}
For $P^{[\fusioncenter]}_{\centeruser,\uav \to \centeruser}  \triangleq \correct^{[\fusioncenter]}_{\centeruser,\uav \to \centeruser} $ and 
    $P^{[\fusioncenter]}_{\centeruser,\centeruser \to \uav} \triangleq 
\correct^{[\fusioncenter]}_{\centeruser,\centeruser \to \uav}$, the probability $\Pr[\correct^{[\fusioncenter]}_{\centeruser_2} ]$ can be derived as
\begin{align}
\Pr\left[\correct^{[\fusioncenter]}_{\centeruser_2} \right] 
    =  1 - P^{[\fusioncenter]}_{\centeruser,\uav \to \centeruser}
    -  P^{[\fusioncenter]}_{\centeruser,\centeruser \to \uav},
\end{align}
where we find that 
    $P^{[\fusioncenter]}_{\centeruser,\uav \to \centeruser}$ and 
    $P^{[\fusioncenter]}_{\centeruser,\centeruser \to \uav}$ can be derived in the same manner as 
    $P^{[\uav]}_{\centeruser \to \edgeuser}$ and 
    $P^{[\uav]}_{\edgeuser \to \centeruser}$, respectively. 
Specifically, we obtain $P^{[\fusioncenter]}_{\centeruser,\uav \to \centeruser}$ and 
    $P^{[\fusioncenter]}_{\centeruser,\centeruser \to \uav}$ by applying the following substitutions in the proof of Lemma \ref{lem_closedForm_P1} and Lemma~\ref{lem:2},
\begin{align}
\begin{array}{cccc}
\Theta_1 := \Theta_{\uav\fusioncenter}, & 
\Theta_2 := \Theta_{\centeruser\fusioncenter}, & 
\hat{\Theta}_3:= \hat{\Theta}_{\uav\fusioncenter}, \\
a_1 := b_2, & \alpha_1 := \beta_2, & A_1 := B_2, & {}\\
a_2 := b_1, & \alpha_2 := \beta_1, & A_2 := B_1, & {\cal A} := \widehat{\mathcal{B}},
\end{array}\!\!
\label{eq_arg_P1P2_to_P5P6}
\end{align}
where $\widehat{\mathcal{B}} \triangleq \frac{B_2 - b_1}{\beta_1}$.
For $P^{[2]}_{\centeruser} \triangleq  \Pr[{\success}^{[2]}_{\centeruser}]$, we have
\begin{align}
P^{[2]}_{\centeruser}
    =  \mathpzc{W}_1(\Theta_{\centeruser\fusioncenter}; 0; b_1)
    =  F_{\centeruser\fusioncenter}^{c}(b_1).
\label{eq_varrho_7_int}
\end{align}

By using the foregoing results, we obtain \eqref{eq_Pout_fc_xC_ana}. This completes the proof.
\end{IEEEproof}

\begin{Theorem} \label{theo_OP_xC1}
Let us denote ${\rm OP}^{[1]}_{\centeruser,{\rm e2e}}$ as the e2e OP of $\centeruser$ in the first phase, which is defined as the probability of failing to decode $x^{[1]}_{\centeruser}$; we have
\begin{align}
{\rm OP}^{[1]}_{\centeruser,{\rm e2e}}
    &=  \Pr[R^{[1]}_\centeruser < R_{\threshold,\centeruser}] 
\label{eq:e2e_OP_xc1_sim} \\
    &=  1 - P^{[2]}_{\centeruser}.
\label{eq_OP_F_xC1}
\end{align}
\end{Theorem}

\begin{IEEEproof}
Plugging \eqref{eq_rate_F_xC2} into \eqref{eq:e2e_OP_xc1_sim}, and after some mathematical steps, we obtain
\begin{align}
{\rm OP}^{[1]}_{\centeruser \to \fusioncenter}
    &= 1- \Pr\left[
        \phi_{\centeruser\fusioncenter} < a_1
    \right] 
    =  1 - P^{[2]}_{\centeruser},
\end{align}
where $P^{[2]}_{\centeruser} \!\triangleq\! \mathpzc{W}_1(\Theta_{\centeruser\fusioncenter}; 0; a_1) = F_{\phi_{\centeruser\fusioncenter}}^c(a_1)$.
This~completes the proof of Theorem \ref{theo_OP_xC1}.
\end{IEEEproof}
\vspace{-0.5cm}
\subsection{Non-adaptive decoding mechanism}

We discuss non-ADM (NADMs) in this subsection.
    In the case of 2-user CDRT-NOMA, there are four alternative non-adaptive decoding orders, denoted as 
    $\vec{d}_{\centeruser\centeruser}$, 
    $\vec{d}_{\edgeuser\centeruser}$,
    $\vec{d}_{\centeruser\edgeuser}$, and
    $\vec{d}_{\edgeuser\edgeuser}$.
{Note that the NADM-aided CDRT-NOMA does not have the same power dependency as the ADM-aided one. The NADM-enabled receivers have predetermined decoding orders due to hardware restrictions, the underlying application, and receivers architectures.}
    
With $\vec{d}_{\centeruser\centeruser}$ and $\vec{d}_{\edgeuser\centeruser}$, $\fusioncenter$ first decodes $\centeruser$'s information signal before decoding $x_{\edgeuser}$, the e2e OPs of $\centeruser$ to decode $x^{[2]}_\centeruser$ and of $\edgeuser$ are obtained as
\begin{align}
{\rm OP}^{[2]}_{\centeruser}( \vec{d}_{ij} )
    &=  1 - \Pr\left[ {\success}^{[\uav]}_{i\to j}, {\success}^{[\fusioncenter]}_{\centeruser\to\uav} \right] 
    -   \Pr\left[ \bar{\success}^{[\uav]}_{i\to j},
        {\success}^{[2]}_{\centeruser} \right], 
\label{eq_e2eOP_xC2_NADM1} \\
{\rm OP}_{\edgeuser}( \vec{d}_{ij} )
    &=   \Pr\left[ \bar{\success}^{[\uav]}_{i\to j} \right] 
    + \Pr\left[ {\success}^{[\uav]}_{i\to j} \right]  
    \Pr\left[ 
        \bar{\success}^{[\fusioncenter]}_{\centeruser\to\uav} \cup \bar{\success}^{[\fusioncenter]}_{\uav}
        \big| \bar{\success}^{[\uav]}_{i\to j}
    \right],
\label{eq_e2eOP_xE_NADM1}
\end{align}
respectively, for $i,j \in \{\centeruser,\edgeuser\}$, $i\ne j$.
    
It is noted that the first NADM is often utilized in cellular networks, here cell-center users are assumed to yield better channel power gains and thus should not be treated as interference in the decoding of $x_{\edgeuser}$ to avoid frequent outages.

With $\vec{d}_{\centeruser\edgeuser}$ and $\vec{d}_{\edgeuser\edgeuser}$, $\fusioncenter$ decodes $\hat{x}_{\edgeuser}$ before decoding $x^{[2]}_{\centeruser}$, the e2e OPs of $\centeruser$ to decode $x^{[2]}_\centeruser$ and of $\edgeuser$ are respectively obtained~as
\begin{align}
{\rm OP}^{[2]}_{\centeruser}( \vec{d}_{ij} )
    &=   \Pr\left[ 
    \bar{\success}^{[\uav]}_{i\to j}, 
    \bar{\success}^{[\fusioncenter]}_{\centeruser\to\uav} \right] 
\nonumber\\
    &\quad+ \Pr\left[ \bar{\success}^{[\uav]}_{i\to j} \right]
    \Pr\left[ \bar{\success}^{[\fusioncenter]}_{\uav\to\centeruser} \cup 
    \bar{\success}^{[\fusioncenter]}_{\centeruser}
    \big| \bar{\success}^{[\uav]}_{i\to j} \right], 
\label{eq_e2eOP_xC2_NADM4} \\
{\rm OP}_{\edgeuser}( \vec{d}_{ij} )
    &=   \Pr\left[ \bar{\success}^{[\uav]}_{i\to j} \right] 
    + \Pr\left[ \bar{\success}^{[\uav]}_{i\to j} \right]  
    \Pr\left[ \bar{\success}^{[\fusioncenter]}_{\uav\to\centeruser}
        |\bar{\success}^{[\uav]}_{i\to j}
    \right].
\label{eq_e2eOP_xE_NADM4}
\end{align}
\subsection{Orthogonal Multiple Access}
Orthogonal Multiple Access (OMA) is a good benchmarking scheme for the NOMA transmission technique. 
    In cooperative networks, if $\fusioncenter$ communicates with $\centeruser$ and $\edgeuser$ in an OMA mode, such as time division multiple access (TDMA), four time slots are needed in total to serve $\centeruser$ and~$\edgeuser$.
Since $\edgeuser$'s information is transmitted over statistically independent channels, i.e., over the A2G and the G2A channels, with the help of a DF-assisted $\uav$, the achievable rate of $\edgeuser$ is determined as 
    ${\cal R}^{\rm OMA}_{\edgeuser} \!=\! \frac{R_{{\sf th},\edgeuser}}{4} \Pr[ \bar{\gamma}_\edgeuser \phi_{\edgeuser\uav} \!>\! \gamma^{\rm OMA}_{{\sf th},\edgeuser}] \Pr[\bar{\gamma}_\uav \phi_{\uav\edgeuser} \!>\! \gamma^{\rm OMA}_{{\sf th},\centeruser} ]$, where $\gamma^{\rm OMA}_{{\sf th},\edgeuser} \triangleq 2^{4 R_{{\sf th},\edgeuser}}-1$. Thus, the achievable rate of $\edgeuser$ is given by
\begin{align}
{\cal R}^{\rm OMA}_{\edgeuser} 
    =   \frac{R_{{\sf th},\edgeuser}}{4}
    F^c_{\phi_{\edgeuser\uav}}\bigg(
        \frac{\gamma^{\rm OMA}_{{\sf th},\edgeuser}}{\bar{\gamma}_{\edgeuser}}
    \bigg)
    F^c_{\phi_{\uav\edgeuser}}
    \bigg(
        \frac{\gamma^{\rm OMA}_{{\sf th},\edgeuser}}{\bar{\gamma}_{\uav}}
    \bigg).
\end{align}

Furthermore, having direct communication with $\fusioncenter$, 
    $\centeruser$ can transmit $x_\centeruser^{[1]}$ and $x_\centeruser^{[2]}$ in two separate time slots, with the corresponding achievable rates of    $\frac{R_{{\sf th},\centeruser}}{4} \Pr[\bar{\gamma}^{[1]}_\centeruser \psi_{\centeruser\fusioncenter} > \gamma^{\rm OMA}_{\thresholdxC}]$ 
    and $\frac{R_{{\sf th},\centeruser}}{4} \Pr[\bar{\gamma}^{[2]}_\centeruser \psi_{\centeruser\fusioncenter} > \gamma^{\rm OMA}_{\thresholdxC}]$, respectively, where $\gamma^{\rm OMA}_{\thresholdxC} \triangleq 2^{4 R_{\thresholdxC}}-1$.
        Hence, the achievable rate of $\centeruser$ is given~by
\begin{align}
{\cal R}^{\rm OMA}_{\centeruser}
=   \frac{R_{{\sf th},\centeruser}}{4}
    \bigg( 
        F^c_{\psi_{\centeruser\fusioncenter}}\bigg(
            \frac{\gamma_{{\sf th},\centeruser}^{\rm OMA}}{\bar{\gamma}_{\centeruser}^{[1]}}
        \bigg)
        + F^c_{\psi_{\centeruser\fusioncenter}}\bigg(
            \frac{\gamma_{{\sf th},\centeruser}^{\rm OMA}}{\bar{\gamma}_{\centeruser}^{[2]}}
        \bigg)
    \bigg).
\end{align}
    
As a result, the achievable sum-rate for OMA transmission protocol is obtained~as
\begin{align}
{\cal R}_\Sigma^{\rm OMA}
    =   {\cal R}^{\rm OMA}_{\centeruser} + {\cal R}^{\rm OMA}_{\edgeuser}.
\end{align}
\vspace{-25pt}
\section{Optimal Power Allocation}

\begin{algorithm}[t]
	\caption{Proposed Iterative Numerical Gradient Descend-based Algorithm for finding ${\cal R}_{\Sigma}^{\ast}$}
    \label{alg_gradient_search}
    \textbf{initialize}: iteration index $\kappa \leftarrow 0$, 
    randomize starting points 
        ${\theta}^{(\kappa)}_1 \in [0,1]$ and 
        ${\theta}^{(\kappa)}_2 \in [0,1]$,
    step size $\tau \leftarrow 0.05$, 
    numerical accuracy 
        $\eta^{(\kappa)}_1, \eta^{(\kappa)}_2 \leftarrow 10^{-4}$,
    and error tolerance 
        $\rho \leftarrow 2.5\times 10^{-3}$. 
    
    \Repeat{
        $\left\Vert (\nabla_{{\boldsymbol{\theta}}} {\cal R}_{\Sigma})({\boldsymbol{\theta}}^{(\kappa)}) \right\Vert_2 < \rho$ 
    }{  \emph{Compute}: ${\cal R}_0^{(\kappa)} \leftarrow {\cal R}_{\Sigma}({\boldsymbol{\theta}}^{(\kappa)})$\;
        \emph{Update}: $\eta^{(\kappa)}_1 \leftarrow \theta_1^{(\kappa)}/10^3$, $\eta^{(\kappa)}_2 \leftarrow \theta_2^{(\kappa)} /10^3$\;
        \emph{Compute}: ${\cal R}_1^{(\kappa)} \leftarrow \frac{{\cal R}_{\Sigma}(\theta_1^{(\kappa)}+\eta^{(\kappa)}_1, \theta_2^{(\kappa)}) - {\cal R}_0^{(\kappa)}}{\eta^{(\kappa)}_1}$\;
        \emph{Compute}: ${\cal R}_2^{(\kappa)} \leftarrow \frac{{\cal R}_{\Sigma}(\theta_1^{(\kappa)}, \theta_2^{(\kappa)} + \eta^{(\kappa)}_2) - {\cal R}_0^{(\kappa)}}{\eta^{(\kappa)}_2}$\;
        \emph{Compute}: 
        $(\nabla_{{\boldsymbol{\theta}}} {\cal R}_{\Sigma})({\boldsymbol{\theta}}^{(\kappa)}) \leftarrow
        \begin{bmatrix}
            {\cal R}_1^{(\kappa)} &
            {\cal R}_2^{(\kappa)}
        \end{bmatrix}^T$\;
        \emph{Update}: 
            ${\boldsymbol{\theta}}^{(k+1)} \leftarrow {\boldsymbol{\theta}}^{(\kappa)} + \tau (\nabla_{{\boldsymbol{\theta}}} {\cal R}_{\Sigma})({\boldsymbol{\theta}}^{(\kappa)})$\;
        \emph{Update}: $k \leftarrow k + 1$\;
    }
    \textbf{output}: ${\boldsymbol{\theta}}^{\ast} = {\boldsymbol{\theta}}^{(\kappa)}$ and 
    ${\cal R}_{\Sigma}^{\ast} = {\cal R}_{\Sigma}({\boldsymbol{\theta}}^{\ast})$.
\end{algorithm}

In this section, we formulate and propose a solution to the system throughput maximization problem. 

\begin{Corollary}
Based on Theorems \ref{theo_OP_xE}-\ref{theo_OP_xC1}, the system throughput can be formulated as follows
\begin{align}
\mathcal{R}_{\Sigma} (\ibf{P})
    &= \frac{ {R}_{\sf th,C} }{ 2 }
        [ 1-{\rm OP}^{[1]}_{\centeruser,{\rm e2e}} (P^{[1]}_{\centeruser}, P_{\edgeuser}, P_{\uav}, P^{[2]}_{\centeruser}) ]
\nonumber\\
    &\quad
    + \frac{ {R}_{\sf th,E} }{ 2 }
        [ 1-{\rm OP}_{\edgeuser,{\rm e2e}} (P^{[1]}_{\centeruser}, P_{\edgeuser}, P_{\uav}, P^{[2]}_{\centeruser}) ]\nonumber\\
    &\quad
    + \frac{ {R}_{\sf th,C} }{ 2 }
        [ 1-{\rm OP}^{[2]}_{\centeruser,{\rm e2e}} (P_{\uav}, P^{[2]}_{\centeruser})],
\label{eq_R_Sigma}
\end{align}
where $\ibf{P} \triangleq [P^{[1]}_{\centeruser}, P_{\edgeuser}, P_{\uav}, P^{[2]}_{\centeruser}]^T$.
\end{Corollary}
First, we formalize the problem as follows
\begin{subequations}
\begin{align}\label{eq30}
    \mathop{\rm maximize}\limits_{\ibf{P}} 
        &\quad \mathcal{R}_{\Sigma}(\ibf{P}) \\
    \text{subject to} 
        &\quad C_1: {P^{[1]}_{\centeruser} + P_{\edgeuser} \le P^{[1]}_{\max}}, \\
    {}  &\quad C_2: {P_{\uav} + P^{[2]}_{\centeruser} \le P^{[2]}_{\max}},
\end{align}
\end{subequations}

Constraints $C_1$ and $C_2$ indicate that the total transmit power of transmitters should not be larger than the power budget in the first and the second time-slots, respectively.
	From \eqref{eq30}, $C_1$ and $C_2$ always hold for the optimal solution.
	Subsequently, the transmit powers can be expressed~as
		${P^{[1]}_{\centeruser} = \theta_1 P^{[1]}_{\max}}$, 
		${P_{\edgeuser} = (1-\theta_1) P^{[1]}_{\max}}$,
		${P^{[1]}_{\centeruser} = \theta_2 P^{[2]}_{\max}}$, and
		${P_{\uav} = (1-\theta_2) P^{[2]}_{\max}}$, where 
		${\theta_{1} \!\triangleq\! \frac{P^{[1]}_{\centeruser}}{P^{[1]}_{\centeruser} + P_{\edgeuser}} \triangleq \frac{P^{[1]}_{\centeruser}}{P^{[1]}_{\max}}}$,
		${\theta_{2} \!\triangleq\! 
		\frac{P^{[2]}_{\centeruser}}{P_{\uav} + P^{[2]}_{\centeruser}} \!=\! \frac{P^{[2]}_{\centeruser}}{P^{[2]}_{\max}}}$.
Accordingly, the optimization problem in \eqref{eq30} can be reformulated~as
\begin{subequations}
\begin{align}\label{eq31}
    \mathop{\rm maximize}\limits_{\ibf{\theta}} 
        &\quad \mathcal{R}_{\Sigma}(\ibf{\theta}) \\
    \text{subject to} 
        &\quad C_1': {0 \le \theta_1 \le 1}, \\
    {}  &\quad C_2': {0 \le \theta_2 \le 1},
\end{align}
\end{subequations}
where $\ibf{\theta} = [\theta_1,\theta_2]^T$.

In order to find the optimal solution to the above problem, derivative-based algorithms, such as the Gradient Descend, can be utilized. 
    In order to perform Gradient Descend, we need to calculate the gradient ${\nabla {\cal R}_{\Sigma}}$ by performing the derivatives of ${{\cal R}_{\Sigma}(\theta_1, \theta_2)}$ with respect to $\theta_1$ and $\theta_2$. 
We thus need to determine
    ${ \frac{\partial}{\partial \theta_1} {\rm OP}_{\edgeuser,{\rm e2e}} }$,
    ${ \frac{\partial}{\partial \theta_1} {\rm OP}^{[1]}_{\centeruser,{\rm e2e}} }$, and
    ${ \frac{\partial}{\partial \theta_1} {\rm OP}^{[2]}_{\centeruser,{\rm e2e}} }$.
Therefore, it will require an excessive number of derivatives to obtain $\nabla {\cal R}_{\Sigma}$. 
    For instance, the derivatives ${ \frac{\partial}{\partial \theta_1} {\rm OP}_{\edgeuser,{\rm e2e}} }$ can be obtained as
\begin{align}
\frac{\partial}{\partial \theta_1} 
{\rm OP}_{\edgeuser,{\rm e2e}}
    &=  - \Pr\left[ \correct_\edgeuser^{[\fusioncenter]} \right] \frac{\partial}{\partial \theta_1} 
    \Pr\left[ \correct^{[\uav]}_{\edgeuser} \right] \nonumber\\
    &\qquad\qquad\quad
        - \Pr\left[ \correct^{[\uav]}_{\edgeuser} \right] 
            \frac{\partial}{\partial \theta_1} 
        \Pr\left[ \correct_\edgeuser^{[\fusioncenter]} \right].
\end{align}

Since $\Pr[ \correct_\edgeuser^{[\fusioncenter]} ]$ is calculated via the events in the second phase, i.e., ${\event}^{[\fusioncenter]}_{\centeruser\to\uav}$, ${\success}^{[\fusioncenter]}_{\centeruser\to\uav}$, ${\success}^{[\fusioncenter]}_{\uav}$ ${\event}^{[\fusioncenter]}_{\uav\to\centeruser}$, and ${\success}^{[\fusioncenter]}_{\uav\to\centeruser}$,
it is independent of $\theta_1$ (i.e., the power allocation in the first phase) thus $\frac{\partial}{\partial \theta_1} {\rm OP}_{\edgeuser,{\rm e2e}}$ can be rewritten as
\begin{align}
\frac{\partial}{\partial \theta_1} {\rm OP}_{\edgeuser,{\rm e2e}}
    =  -\Pr\left[ \correct_\edgeuser^{[\fusioncenter]} \right]
    \frac{\partial}{\partial \theta_1} 
    \Pr\left[ \correct^{[\uav]}_{\edgeuser} \right].
\end{align}

Calculating the derivative of ${\Pr[ \correct^{[\uav]}_{\edgeuser} ]}$ requires the derivatives of $\rho_{1,1}$, $\rho_{1,2}$, $\rho_{1,3}$, $\rho_{1,4}$, $\rho_{2,1}$, $\rho_{2,2}$, and $\rho_{2,3}$, which proves to be excessive.
    To overcome this mundane, we propose a Gradient Descend-inspired algorithm in Algorithm \ref{alg_gradient_search}. 
In this Algorithm, we perform traditional Gradient descend to find ${\boldsymbol{\theta}}^\ast$ and ${\cal R}_{\Sigma}({\boldsymbol{\theta}}^\ast)$, which represent the optimal UL power allocation and the optimal throughput, respectively.
    During each $k$-th iteration, instead of analytically calculating the Gradient of ${\cal R}_{\Sigma}$, we numerically evaluate $\nabla{{\cal R}_{\Sigma}}$, as in Step 3 of Algorithm \ref{alg_gradient_search}, by adopting the following identity
\begin{align}
    \frac{ f(x+\eta) - f(x) }{ \eta } 
    \mathop{\to}\limits^{\eta \to 0} \frac{\partial f }{\partial x}(x),
\end{align}
where ${\eta} > 0$ represents the numerical accuracy satisfying
\begin{align}
    \big| f(x+\eta) - f(x) \big| > 0.
\end{align}

Then, the power allocation is updated for the next iteration as in Step 4, where ${{\boldsymbol{\theta}}^{(\kappa)} = \big[ \theta_1^{(\kappa)}, \theta_2^{(\kappa)} \big]^T}$.
    The algorithm iterates until the stopping criterion in Step 10 is satisfied.
When $\eta_1^{(\kappa)}$ or $\eta_2^{(\kappa)}$ is inappropriately set so that both ${\cal R}_1^{(\kappa)}$ and ${\cal R}_2^{(\kappa)}$ return zero, the algorithm terminates early, resulting in erroneous, suboptimal power allocation.
It should be noted that the proposed optimization algorithm does not rely on the knowledge of decoding events at $\uav$ and $\fusioncenter$. Instead, it is based on analytical results of the sum-rate, which only requires known parameters, such as the location of each node, the channel severity factors between nodes, and the instantaneous target spectral efficiency.

\section{Performance Analysis of a Nonstationary UAV}

Unlike previous sections, we consider that the locations of the UEs follow the random waypoint mobility (RWM) model \cite{HyytiaTMC2006}. 
    In particular, the movement of the UEs are restricted within 2-Dimensional (2D) circular area ${\area}_{\sf X}$ with the radius ${R}$ and the center is located at ${\bf p}_{{\sf X},0}$, ${\sf X} \!\in\! \{ \centeruser,\edgeuser \}$.
    
Let us denote $\vec{p}_{{\sf X},i}$ as the coordinates of the $i$-th waypoint, $i = 1,2,\dots$, that the UEs choose in the $i$-th movement period, the movement trace of the UEs can be modeled as discrete-time stochastic processes 
$\{ \vec{p}_{{\sf X},i} \}_{i\in \mathbb{N}} 
    \!=\! \{ \vec{p}_{{\sf X},1},
             \vec{p}_{{\sf X},2},\dots \}
$.
The waypoints $\vec{p}_{{\sf X},i} \!=\! [ X_{{\sf X},i}, Y_{{\sf X},i}, Z_{\sf X} ]$ are statistically i.i.d. RVs and are uniformly distributed over $\area_{\sf X}$.
    The movement of each UE from the initial waypoint ${ \vec{p}_{{\sf X},i} }$ to the next waypoint $\vec{p}_{{\sf X},i+1}$ is described as follows.
First, each UE chooses a new velocity ${ v_{{\sf X},i} \in [v_{{\sf X},\min}, v_{{\sf X},\max}] }$. 
    Then, the UE moves along the line segment from ${ \vec{p}_{{\sf X},i} }$ to ${ \vec{p}_{{\sf X},i+1} }$ in the direction $\psi_{{\sf X},i} \!\triangleq\! {\arctan} \big( \frac{Y_{{\sf X},i+1} - Y_{{\sf X},i}}{X_{{\sf X},i+1} - X_{{\sf X},i}} \big)$ with the chosen velocity $v_{{\sf X},i}$. 
    
Let $\Delta t$ be the time interval between each two consecutive captured locations, the coordinates of the UEs at ${(t \!+\! \Delta t)}$ is
    $ {{\bf p}_{{\sf X}}[t+\Delta t] = {\bf p}_{{\sf X}}[t] + {\bf v}_{{\sf X},i} \Delta t} $,
where ${{\bf v}_{{\sf X},i} \!\triangleq\! [ v_{{\sf X},i} \cos\psi_{{\sf X},i}, v_{{\sf X},i} \sin\psi_{{\sf X},i}, 0 ]}$ is the motion vector of ${\sf X}$.
After reaching ${\vec{p}_{{\sf X},i+1}}$, each UE has a probability of ${\omega_{{\sf X},i}}$ to remain stationary for a time of $T_{{\sf X}, i}$ before moving to the next waypoint.
    In this research, we assume that ${{\omega_{\sf X}} = 0}$ and $v_{{\sf X},i}$ is uniformly distributed within the interval $[v_{{\sf X},\min},v_{{\sf X},\max}]$.

\subsubsection{Statistical Characteristic of $\centeruser$-$\fusioncenter$ Distance}

The distance between $\centeruser$ and $\fusioncenter$ corresponds to the distance between a RWM model-based point and its reference point.
    Therefore, by following the analysis in \cite{HyytiaTMC2006}, the PDF of ${d_{\centeruser\fusioncenter}}$ is formulated~as
\begin{align}
\label{eq_dCF_form}
    f_{d_{\centeruser\fusioncenter}}(r) 
        =   2\pi r  
        \frac{\eta(R; r)}{ C_\eta(R) },
    ~r < R,
\end{align}
where
$\eta(R; r) \!\triangleq\! 2(R^2 \!-\! r^2) \int_0^{\pi}{ \sqrt{R^2 \!-\! r^2\cos^2\phi} {\rm d}\phi }$ and  $C_\eta(R) \!\triangleq\! \frac{128 \pi R^5}{45}$.
It is noted that $\eta(R; r)$ is a function of the incomplete elliptic integral of the second kind and cannot be derived in closed-form expressions.

\begin{Lemma}
\label{lem:dCF}
The PDF of the distance from $\centeruser$ to $\fusioncenter$
can be formulated as
\begin{align}
\label{eq_dCF_approx}
    f_{d_{\centeruser\fusioncenter}}(r) 
        \approxeq   2\pi r 
        \frac{ P_3(R; r) }{ C_P(R) },
    ~r < R,
\end{align}
where $P_3(R; r) \!=\! \frac{ 3 (R^2 \!-\! r^2) }{ 257\pi} 
        (189 \!-\! 44\frac{r^2}{R^2} \!-\! 18\frac{r^4}{R^4})$,
    $C_P(R) \!=\! R^4$.
\end{Lemma}

\subsubsection{Statistical Characteristic of $\edgeuser$-$\fusioncenter$ Distance}
    The following Lemma provides the PDF of $d_{\edgeuser\fusioncenter}$.

\begin{Lemma}
\label{lem:dEF}
The PDF of the distance from $\edgeuser$ to $\fusioncenter$, given that ${X_{\edgeuser,0} \!=\! D_0 \cos\varphi_{\edgeuser\fusioncenter,0}}$ and 
    ${Y_{\edgeuser,0} \!=\! D_0 \sin\varphi_{\edgeuser\fusioncenter,0}}$, is obtained~as
\begin{align}
f_{d_{\edgeuser\fusioncenter}}(r)
    &= 
    \left\{
    \begin{array}{ll}
    \displaystyle
        f_1(D_0,R; r), & \hfill |D_0-R| \le r \le D_0 + R, \vspace{1pt}\\
    \displaystyle
        f_2(D_0,R; r), & \hfill r \le R-D_0,
        D_0 < R,
    \end{array}
    \right. \nonumber\\
&\quad
\text{for} ~ (D_0-R)^+ < r < D_0+ R,
\end{align}
where $f_1(D,R; r)$ and $f_2(D,R; r)$ are given by \eqref{eq_f1} and \eqref{eq_f2} at top of the next page, respectively,
\begin{table*}
\begin{align}
    f_1(D,R;r) 
    &=
    \frac{6 r}{257\pi R^8}
    \bigg[
    -\sqrt{ (r^2-(D-R)^2)((D+R)^2-r^2) } (C_1(D,R) + 33 r^4 + C_2 (D,R) r^2) \nonumber\\
    &\qquad\qquad\qquad\qquad\qquad
    + ( C_3(D,R) + 18 r^6 + C_4(D,R) r^4 + C_5(D,R) r^2 ) 
    \cos^{-1}\left( \frac{D^2 + r^2 - R^2}{2 D r} \right)
    \bigg], \label{eq_f1} \\
    f_2(D,R;r) 
    &= 
    \frac{6 r}{257 R^8}
    [ C_3(D,R) + 18 r^6 + C_4(D,R) r^4 + C_5(D,R) r^2 ], \label{eq_f2}
\end{align}
\hrulefill
\end{table*}
in which
\begin{subequations}
\begin{align}
    C_1(D,R) &= 33 D^4 + 54 D^2 R^2 - 214 R^4, \\
    C_2(D,R) &= 114 D^2 + 54 R^2, \\
    C_3(D,R) &= 18 D^6 + 26 D^4 R^2 - 233 D^2 R^4 + 189 R^6, \\
    C_4(D,R) &= 162 D^2 + 26 R^2, \\
    C_5(D,R) &= 162 D^4 + 104 D^2 R^2 - 233 R^4.
\end{align}
\end{subequations}
\end{Lemma}
\begin{IEEEproof}
See Appendix \ref{apx:D}.
\end{IEEEproof}

\subsubsection{Statistical Characteristic of $\uav$-$\edgeuser$ Distance}

To represent the correlation between the movement of the UAV and the movement of $\edgeuser$, we propose that the movement of $\uav$ follows the Reference Point Group Mobility (RPGM) model, in which the $\edgeuser$ serves as the group center \cite{Hong1999}. 
    Thus, the behavior of the UAV's motion, including its trajectory and speed, is directly influenced by $\edgeuser$'s motion.
At any given time $t$, we have ${\big| {\bf p}_{\uav}[t]-{\bf p}_{\edgeuser}[t] \big| \le R_d}$, where ${\bf p}_{\uav}[t]$ is the captured position of $\uav$ at time $t$, $R_d$ is the maximum deviation of $\uav$'s position to $\edgeuser$'s position.
    We denote ${\bf r}_{\uav,i}$ and ${\bf p}_{\uav,i}$ as the coordinates of the $i$-th reference point and the $i$-th position of $\uav$.
Generalization-wise, $\uav$ follows the movement of $\edgeuser$ with an addition of a small random deviation from $\edgeuser$'s motion vector. When the random factors of the RPGM model are ignored, $\uav$ moves along a trajectory that is designed based on $\edgeuser$'s motion.
    The UAV's movement in the RPGM model is described as follows:
\begin{itemize}
    \item First, the reference point ${\bf r}_{\uav,i}$ moves to the reference point ${\bf r}_{\uav,i+1}$ with the group motion vector ${\bf v}_{\edgeuser,i}$.
    \item Second, a new UAV's location ${\bf p}_{\uav,i+1}$ is generated by adding a random motion vector ${\bf v}_{i+i}$ deviated from the new reference point. We consider $|{\bf v}_{i+1}| \in [v_{\uav,\min},v_{\uav,\max}]$, where $v_{\uav,\min}$ and $v_{\uav,\max}$ are the minimum and the maximum deviation in the $\uav$'s velocity compared to $\edgeuser$.
    \item Third, the UAV moves from ${\bf r}_{\uav,i}$ to ${\bf r}_{\uav,i+i}$ with a motion vector satisfying ${\bf v}_{\uav, i} + {\bf v}_{i} \!=\! {\bf v}_{\edgeuser,i} + {\bf v}_{i+1}$. 
\end{itemize}

It is noted that the position of $\uav$ captured at $(t\!+\!\Delta)$ is formulated as
$\vec{p}_{\uav}[t+\Delta] = \vec{p}_{\uav}[t]+\vec{p}_{\edgeuser}[t]+\vec{v}_{{\sf U},i} \Delta t$. 
    The PDF of the distance between the UAV and the $\UEE$ is given by the following Lemma.
\begin{Lemma}
\label{lem:dUE}
The PDF of the distance from $\edgeuser$ to $\uav$, hovering at an altitude $H$ from the ground, can be formulated~as
\begin{align}
f_{d_{\edgeuser\uav}}(r) 
    \approxeq
    2\pi r \frac{ P_3(R_d; \sqrt{r^2-H^2}) }{ C_P(R_d) },
    H < r < \textstyle{\sqrt{R_d^2+H^2}}.
\label{eq:dUE}
\end{align}
\end{Lemma}
\vspace{-5pt}

\begin{IEEEproof}
Let $\underline{\uav}$ be the projection of $\uav$ on the ground, then the PDF of the distance between $\underline{\uav}$ and $\edgeuser$ is given by
\begin{equation}
f_{d_{\edgeuser\underline{\uav}}}(r) 
    \approxeq   
    2\pi r
    \frac{ P_3(R_d; r) }{ C_P(R_d) },~0 < r < R_d.
\end{equation}
Since $(d_{\edgeuser\uav})^2 \!=\! {(d_{\edgeuser\underline{\uav}})^2 + H^2}$, the PDF of $d_{\edgeuser\uav}$ is derived~as
{\allowdisplaybreaks
\begin{align}
&f_{d_{\edgeuser\uav}}(r)
    =   \frac{\rm d}{{\rm d} r}
    \Pr[d_{\edgeuser\uav} \!<\! r]
\label{eq_dUE_derivation_1} \\
    &\quad=  
    \frac{\rm d}{{\rm d} r}
    \Pr[d_{\edgeuser\underline{\uav}} \!<\! {\textstyle \sqrt{r^2-H^2}}, r > H, d_{\edgeuser\underline{\uav}} < 2 R_d]\! 
\label{eq_dUE_derivation_2} \\
    &\quad=  \frac{\rm d}{{\rm d} r}
    \Pr[d_{\edgeuser\underline{\uav}} \!<\! {\textstyle \sqrt{r^2-H^2}},  \sqrt{(2R)^2+H^2} \!>\! r \!>\! H]\! \nonumber\\
    &\quad\quad\quad+   \frac{\rm d}{{\rm d} r}
    \Pr[d_{\edgeuser\underline{\uav}} \!<\! 2R_d, {\textstyle \sqrt{(2R_d)^2+H^2}} \!<\! r]\! 
\label{eq_dUE_derivation_3} \\
    &\quad=  \frac{\rm d}{{\rm d} r} F_{d_{\edgeuser\underline{\uav}} }( {\textstyle \sqrt{r^2-H^2})},
    ~{\textstyle \sqrt{(2R_d)^2+H^2} } \!>\! r \!>\! H.
\label{eq_dUE_derivation_4}
\end{align}
}
Performing the above derivatives gives \eqref{eq:dUE}. This completes the proof of Lemma \ref{lem:dUE}.
\end{IEEEproof}

\subsubsection{Statistical Characteristic of $\uav$-$\fusioncenter$ Distance}
The distance between $\uav$ and $\fusioncenter$ corresponds to the distance between a RPGM model-based point and the origin. 
The following Lemma provides the PDF of $d_{\uav\fusioncenter}$.

\begin{Lemma}
\label{lem:dUF}
The PDF of the distance between $\uav$ and $\fusioncenter$ can be formulated as
\begin{gather} \label{eq:dUF_final}
f_{d_{\uav\fusioncenter}}(r) 
    \approxeq \frac{r}{\sqrt{r^2-H^2}} 
    f_{d_{\underline{\uav}\fusioncenter}}({\textstyle \sqrt{r^2-H^2}}), \\
    \text{for } 
    {\textstyle \sqrt{ (D_0-R)^2+H^2 } < r < \sqrt{ (D_0 + R)^2 + H^2} }. \nonumber
\end{gather}
\end{Lemma}
\begin{IEEEproof}
When $R_d \ll R$, the PDF of the distance from $\underline{\uav}$ to $\fusioncenter$ can be approximated as $f_{d_{\underline{\uav}\fusioncenter}}(r) \approxeq f_{d_{\edgeuser\fusioncenter}}(r)$, for $(D_0-R)^+ \!<\! r \!<\! D_0+ R$.
Since $(d_{\uav\fusioncenter})^2 \!=\! (d_{\edgeuser\underline{\uav}})^2+H^2$, by applying the derivations in \eqref{eq_dUE_derivation_1}-\eqref{eq_dUE_derivation_4}, we obtain \eqref{eq:dUF_final}. This completes the proof of Lemma~\ref{lem:dUF}.
\end{IEEEproof}

\subsubsection{Statistical Characteristic of $\uav$-$\centeruser$ Distance}
The distance between $\uav$ and $\centeruser$ corresponds to the distance between a RPGM model-based point and the RWM-based point. 
    Let us denote $\underline{\uav}_0$ whose position is obtained as ${\bf p}_{\underline{\uav}_0} = {\bf p}_{\underline{\uav}} + {\bf p}_{\edgeuser,0}$, the PDF of the distance from $\centeruser$ to $\underline{\uav}_0$ is obtained as
\begin{align}
\label{eq:dCU0}
f_{d_{\centeruser\underline{\uav}_0}}(r) 
    \approxeq   \frac{r P_0(r/R)}{R^2},~0 < r < 2 R,
\end{align}
where
\begin{align}
P_0(r) 
    =  \frac{45}{64}
    \bigg\{
        \int_{|1-r|}^{1} 
        \!\!\!\!\!\!\!\!\! 
        { \eta(\tau) f_{1}(\tau,1;r){\rm d} {\tau} }
    \!+\! \int_0^{(1 - r)^+}
        \!\!\!\!\!\!\!\!\!\!\!\!\!\!\!\!
        \eta(\tau) f_{2}(\tau,1;r){\rm d} {\tau}
    \bigg\}.
\end{align}

\begin{Lemma}
\label{lem:dCU}
Based on \eqref{eq:dCU0}, the PDF of the distance from $\centeruser$ to $\uav$ can be accurately formulated by the two following cases:
\begin{align}
&f_{d_{\centeruser\uav}}(r)
    =   \frac{r}{2 \pi R^2} 
\nonumber\\
    &\times
    \left\{
    \begin{array}{lc}
    \displaystyle
        \int_{T_0- \nu_c (r)}^{T_0+ \nu_c (r)} 
        \tilde{P}_0( {\textstyle \sqrt{ r^2-H^2 }},t ) {\rm d} t, \\ 
        \text{ if } {\textstyle \sqrt{(D_0-2R)^2 +H^2}} \le r 
        \hfill \le {\textstyle \sqrt{(D_0+2R)^2 +H^2}}, \\
    \displaystyle
        \int_{0}^{2\pi} 
        \tilde{P}_0( {\textstyle \sqrt{ r^2-H^2 }},t ) {\rm d} t, \\
        \hfill \text{ if } {\textstyle H} \!\le\! r \!\le\! {\textstyle \sqrt{\max(2R-D_0,0)^2+H^2}},
    \end{array}
    \right.
\end{align}
where $T_0 = \tan^{-1}( \frac{Y_{\edgeuser,0}}{X_{\edgeuser,0}})$, 
    $\nu_c (r) \triangleq \cos^{-1}\big( \frac{r^2+D_0^2-4R^2}{2 r D_0} \big)$, and
\begin{align}
\tilde{P}_0(r,t) \triangleq 
    P_0\bigg( \frac{\sqrt{ r^2+D_0^2-2 D_0 r\cos(t-T_0) }}{R} \bigg).
\end{align}
\end{Lemma}

Using the approach in \cite{HyytiaTMC2006}, an approximation of $P_0(r)$ can be obtained as
\begin{align}
P_0(r) &\approxeq C_0 r (4 - r^2)^4 (376 - 101 r^2 \nonumber\\ 
    &\qquad\qquad\qquad\quad
    + 48 r^4 - 11 r^6 + r^8),~r < 2.
\end{align}

Let $\overline{\cal R}_{\Sigma} $ be the expectation of ${\cal R}_{\Sigma}$ over the captured trajectory ${\boldsymbol{\mathfrak{\rho}}}$, where 
    $[{\boldsymbol{\mathfrak{\rho}}}]_i = [ \vec{p}_{\edgeuser,i}, \vec{p}_{\centeruser,i}, \vec{p}_{\uav,i} ]$, ${i = 0,1,2,\dots,\infty}$, thus $\overline{\cal R}_{\Sigma} $ can be derived as
\begin{align}
\overline{\cal R}_{\Sigma}
    &=  \mathbb{E}\left\{ {\cal R}_{\Sigma}({\boldsymbol{\mathfrak{\rho}}}_c) \right\}
    =  \mathbb{E}\left\{ {\cal R}_{\Sigma}(d_{\uav\fusioncenter},d_{\edgeuser\uav},d_{\centeruser\fusioncenter},d_{\centeruser\uav}) \right\}
    \\
    &\approxeq 
    \int_{\sqrt{(D_0-R)^2 + H^2}}^{\sqrt{(D_0+R)^2 + H^2}}
    \int_{H}^{\sqrt{R_d^2+H^2}-H^2}
    \int_{0}^{R}
    \int_{H}^{\sqrt{(D_0+2 R)^2 + H^2}}
\nonumber\\
    &\quad\times
        f_{d_{\uav\fusioncenter}}(x)
        f_{d_{\edgeuser\uav}}(y)
        f_{d_{\centeruser\fusioncenter}}(z)
        f_{d_{\centeruser\uav}}(t)
        {\cal R}_{\Sigma}(x,y,z,t) 
        {\rm d} x  {\rm d} y {\rm d} z {\rm d} t,
\label{eq:corollary2}
\end{align}
where ${\cal R}_{\Sigma}( d_{\uav\fusioncenter},d_{\edgeuser\uav},d_{\centeruser\fusioncenter},d_{\centeruser\uav} )$ is the throughput in terms of $d_{\uav\fusioncenter}$, $d_{\edgeuser\uav}$, $d_{\centeruser\fusioncenter}$, and $d_{\centeruser\uav}$.
It is intractable to derive the exact closed-form expressions of \eqref{eq:corollary2}, thus we make use of the Gaussian Chebyshev quadrature \cite{hildebrand1987} to further approximate $\overline{\cal R}_{\Sigma}$, as resulted in the following corollary.

\begin{Corollary}
The approximate closed-form expression of $\overline{\mathcal{R}}_{\Sigma}$ is obtained as
\begin{align}
\overline{\mathcal{R}}_{\Sigma}
    &\approxeq
    \frac{C^-_{\uav\fusioncenter}}{2}
    \sum_{p=1}^{N_{\uav\fusioncenter}}
    \frac{\pi}{N_{\uav\fusioncenter}}
    {\textstyle \sqrt{1-(\phi^{(p)}_{\uav\fusioncenter})^2}}
    f_{d_{\uav\fusioncenter}}(x_p) 
\label{eq:R_Sigma_GC} \\
    &\quad\times
    \frac{C^-_{\edgeuser\uav}}{2}
    \sum_{q=1}^{N_{\edgeuser\uav}}
    \frac{\pi}{N_{\edgeuser\uav}}
    {\textstyle \sqrt{1-(\phi^{(q)}_{\edgeuser\uav})^2}}
    f_{d_{\edgeuser\uav}}(y_q) 
\nonumber\\
    &\quad\times
    \frac{R}{2}
    \sum_{r=1}^{N_{\centeruser\fusioncenter}}
    \frac{\pi}{N_{\centeruser\fusioncenter}}
    {\textstyle \sqrt{1-(\phi^{(r)}_{\centeruser\fusioncenter})^2}} f_{d_{\centeruser\fusioncenter}}(z_r)
\nonumber\\
    &\quad\times
    \frac{C^-_{\centeruser\uav}}{2}
    \sum_{s=1}^{N_{\centeruser\uav}}
    \frac{\pi}{N_{\centeruser\uav}}
    {\textstyle \sqrt{1-(\phi^{(s)}_{\centeruser\uav})^2}}
    f_{d_{\centeruser\uav}}(t_s)
    {\cal R}_{\Sigma}(x_p,y_q,z_r,t_s),
\nonumber
\end{align}
where ${\phi^{(i)}_{\sf XY} = \cos\big( \frac{2 i-1}{2 N_{\sf XY}} \pi \big),~{C^\pm_{\centeruser\uav} \!=\! {\textstyle \sqrt{(D_0+ 2R)^2 + H^2}} \pm H}}$,
    ${C^\pm_{\uav\fusioncenter} \!=\!  {\textstyle \sqrt{(D_0\!+\!R)^2 \!+\! H^2}} \pm {\textstyle \sqrt{(D_0-R)^2 + H^2}}}$,
    ${C^\pm_{\edgeuser\uav} \!=\! {\textstyle \sqrt{R_d^2 \!+\! H^2} \pm H}}$,
    $x_p \!=\!  \frac{C^+_{\uav\fusioncenter}}{2}
    \!+\! \frac{C^-_{\uav\fusioncenter}}{2} \phi^{(p)}_{\uav\fusioncenter}$,
    $y_q \!=\! \frac{C^+_{\edgeuser\uav}}{2} \!+\! \frac{C^-_{\edgeuser\uav}}{2} \phi^{(q)}_{\edgeuser\uav}$,
    $z_r \!=\! \frac{R}{2} \big( 1 \!+\! \phi^{(r)}_{\centeruser\fusioncenter} \big)$, 
    and
    $t_s \!=\! \frac{C^+_{\centeruser\uav}}{2} \!+\! \frac{C^-_{\centeruser\uav}}{2} \phi^{(s)}_{\centeruser\uav}$.
\end{Corollary}

In addition, $\overline{\cal R}_{\Sigma}$ can also be approximated as
\begin{align}
\label{eq_avgRSigma_pos}
\overline{\cal R}_{\Sigma} 
    = \lim\limits_{N\to\infty}
    \frac{1}{N}
    \sum_{i=1}^{N}
    {\cal R}_{\Sigma}( [{\vec{p}}]_i )
    \approx 
    \frac{1}{N}
    \sum_{i=1}^{N}
    {\cal R}_{\Sigma}( [{\vec{p}}]_i ),
\end{align}
where ${\cal R}_{\Sigma}( [{\vec{p}}]_i )$ is the throughput measured at  $[{\vec{p}}]_i$ and $N$ is the number of observations. The accuracy of this approximation is later presented in the next Section.

\vspace{-5pt}
\section{Numerical Results}
\vspace{-2.5pt}

\begin{figure}
    \centering
    \includegraphics[width = 0.8\linewidth]{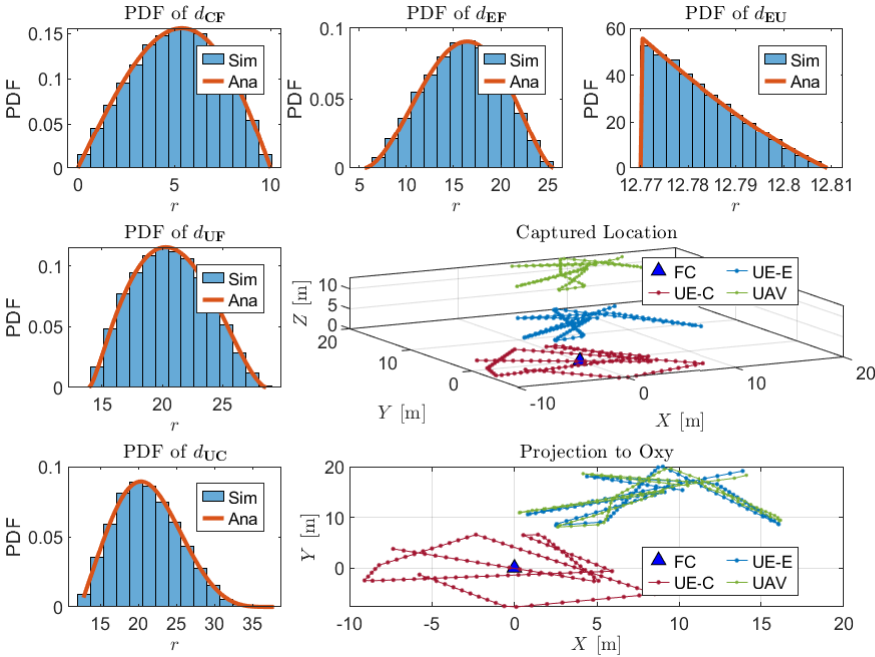}
    \caption{{Distance distribution and top-down view of 125 captured locations of $\uav$ following RPGM model and of $\edgeuser$ and $\centeruser$ following RWM model, where FC is located at the origin}.}
    \label{fig:2}
\vspace{-0.3cm}
\end{figure}

\begin{figure}[t]
    \centering
    \includegraphics[width=.7\linewidth]{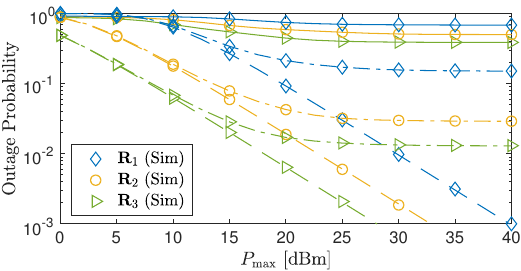}
    \caption{Impact of the maximum power budget using derived analytical formulations on ${\rm OP}_{\edgeuser,{\rm e2e}}$ (solid), 
    ${\rm OP}^{[1]}_{\centeruser,{\rm e2e}}$ (dashed), ${\rm OP}^{[2]}_{\centeruser,{\rm e2e}}$ (dot-dash).}
\label{eq_OP_vs_Pmax}
\end{figure}

\begin{figure}[t]
    \centering
    \includegraphics[width=.75\linewidth]{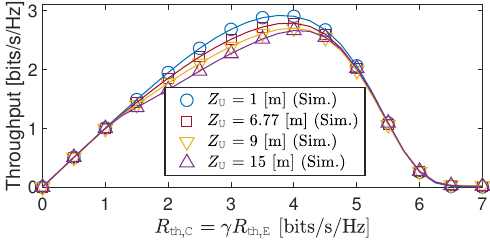}
    \caption{Throughput versus the target SE and $\uav$'s altitude (i.e., $Z_{\uav}$).}
    \label{fig_throughput_vs_specEff}
\vspace{-0.3cm}
\end{figure}

\begin{figure}
    \centering
    \includegraphics[width = 0.7\linewidth]{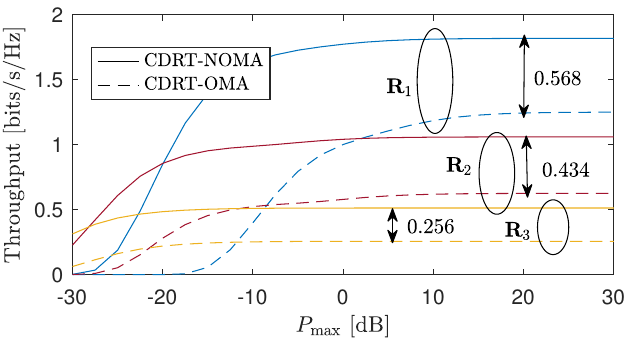}
    \caption{{Throughput of ADM-enabled CDRT-NOMA and CDRT-OMA versus $P_{\max}$ with different target spectral efficiency. \label{fig:3p1_1}}}
\vspace{-0.5cm}
\end{figure}

\begin{figure}[t]
	\centering
	\subfloat[]{%
		\includegraphics[width=.35\linewidth]{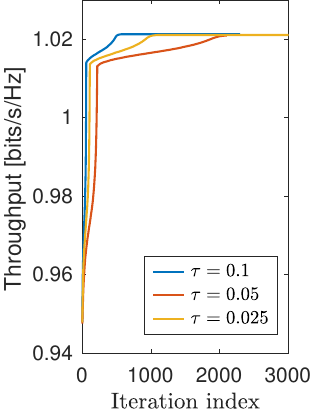}
		\label{fig_NGD_convergence}} 
	\subfloat[]{%
		\includegraphics[width=.35\linewidth]{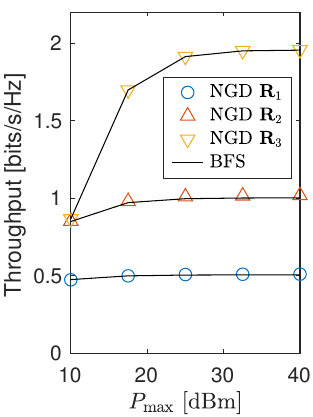}
		\label{fig_BFS_NGD_vs_location}} 
	\caption{(a) The convergence speed of the numerical gradient descend-based (NGD) algorithm and (b) performance comparison between the brute-force search (BFS) and NGD algorithms.}
	\label{} 
\end{figure}

\begin{figure}[htp]
    \centering
    \includegraphics[width=0.7\linewidth]{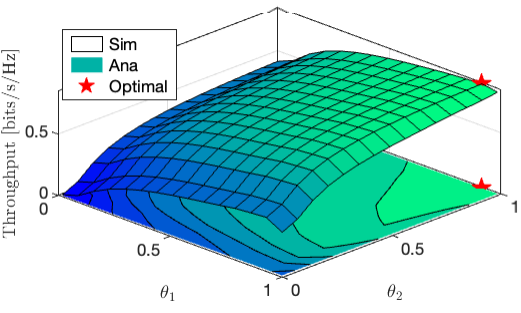}
    \caption{Throughput at the initial location of $\uav$ as a function of $\theta_1$ and $\theta_2$.}
    \label{fig:7}
\end{figure}

\begin{figure}[!h]
    \centering
    \includegraphics[width = 0.75\linewidth]{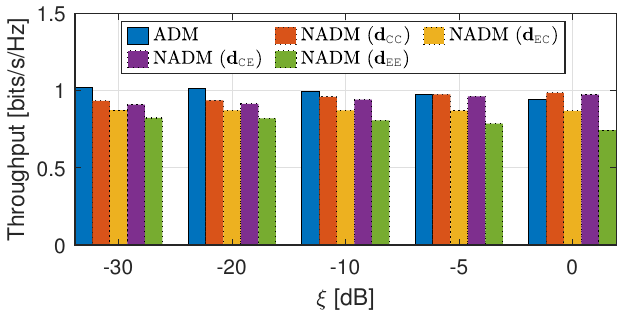}
    \caption{Throughput of ADM and NADM versus the RI level at $\uav$ and $\fusioncenter$, when $\xi_\uav = \xi_\fusioncenter = \xi$. 
    \label{fig_throughput_vs_RILevel} }
\end{figure}

\begin{figure}[htp]
    \centering
    \includegraphics[width=0.7\linewidth]{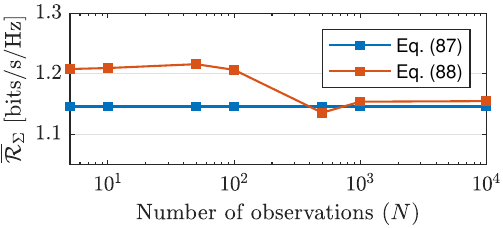}
    \caption{{Accuracy of the average throughput in \eqref{eq_avgRSigma_pos} compared to \eqref{eq:R_Sigma_GC}.}}
\label{eq_OP_vs_angle}
\end{figure}

\begin{figure}[!h]
    \centering
    \subfloat[\label{fig_insRSigma_vs_Locationa}]{%
       \includegraphics[width = 0.45\linewidth]{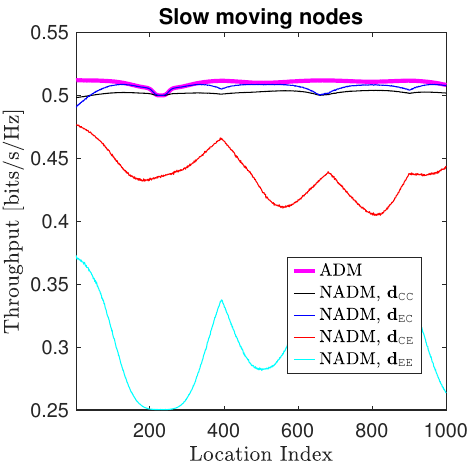}}
    \quad
    \subfloat[\label{fig_insRSigma_vs_Locationb}]{%
       \includegraphics[width = 0.45\linewidth]{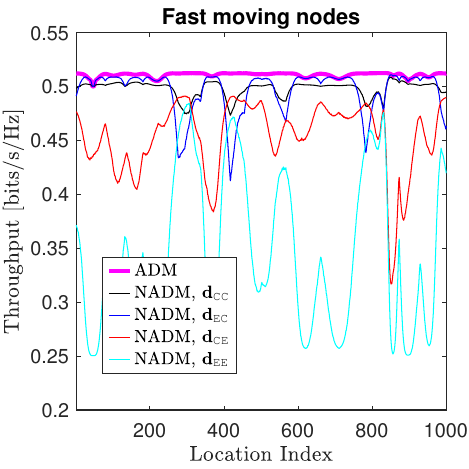}}
    \caption{{ADMs versus NADM over $10^3$ observations for a) slow moving nodes, and b) fast moving nodes, where the residual level is $-30$ [dB].\label{fig_insRSigma_vs_Location}}}
\end{figure}

    This section includes numerical results to validate the correctness of the analysis provided in previous sections. 
    {We consider 3GPP Urban Micro (UMi) path loss model for LoS communication, ${\pl}_{\centeruser\fusioncenter} = G_{\centeruser} + G_{\fusioncenter} + 22.7 + 26\log(f_c)$ [dB] and $\epsilon = 3.67$ \cite[Table B.1.2.1-1]{NOMA3GPP}. The environmental parameters are $\alpha_1 = 12.08$ and $\alpha_2 = 0.11$ for dense urban areas \cite{HouraniWCL2014}.}
Furthermore, with the obtained results, we provide additional insights on the performance of the proposed UAV-aided double uplink system.
    In what follows, unless otherwise specified, the simulation settings are those that are provided in Table~\ref{table_parameters}.
{From Fig. \ref{eq_OP_vs_Pmax}-\ref{fig_throughput_vs_RILevel}, we consider the location of $\centeruser$, $\edgeuser$, and $\uav$ are fixed, where $\vec{p}_\centeruser = (-1.96,7.33,0)$, $\vec{p}_\edgeuser = (-13.49,-18.85,0.23)$, and $\vec{p}_\uav = (-6.66,-7.62,6.77)$, respectively, where $\sigma^2 = -71$ dBm. 
Observations of throughput in a nonstationary environment are shown in Fig. \ref{eq_OP_vs_angle} and Fig. \ref{fig_insRSigma_vs_Location}, where $\sigma^2 = -104$~dBm. In this context, a realization of the locations of $\uav$, $\edgeuser$, and $\centeruser$ are shown in Fig. \ref{fig:2}.}

\begin{table}[t]
	\centering
	\caption{Simulation Parameters}{
	\begin{tabularx}{.9\linewidth}{X c }
		\Xhline{2\arrayrulewidth}
         \textbf{Parameter} & \textbf{Value} \\
		\Xhline{2\arrayrulewidth}
		{Maximum network dimension [m]} & {50} \\
		{Target spectral efficiency of $\centeruser$ [bits/s/Hz]} & {1.0} \\
		{Residual interference level [dB]} & {-10} \\
		{Target spectral efficiency of $\edgeuser$ [bits/s/Hz]} & {0.05} \\
		{Transmit Antenna gains [dBi]} & {0} \\
		{Carrier frequency, $f_c$ [GHz]} & {3} \\
		{Receive Antenna gains [dBi]} & {0} \\
		{LoS Link attenuation, $\eta$, [dB]} & {1.6} \\
		{NLoS Link attenuation, $\eta$, [dB]} & {23} \\
		{NLoS Link attenuation, $\bar{\eta}$, [dB]} & {23} \\
		{LoS fading severity, $m_{\centeruser\fusioncenter}$, $m_{\centeruser\uav}$, $m_{\edgeuser\uav}$, $m_{\uav\fusioncenter}$} & {5, 3, 1, 5} \\
            Minimum velocity of UEs, $v_{\centeruser,\min}=v_{\edgeuser,\min}$, [m/s] & $0.1$ \\
            Maximum velocity of UEs, $v_{\centeruser,\max}=v_{\edgeuser,\max}$, [m/s] & $0.2$ \\
            Min. velocity deviation, $v_{\uav,\min}=v_{\uav,\min}$, [m/s] & $0.01$ \\
            Max. velocity deviation, $v_{\uav,\max}=v_{\uav,\max}$, [m/s] & $0.05$ \\
		\Xhline{2\arrayrulewidth}
	\end{tabularx} }
	\label{table_parameters}
\end{table}

\textit{Impact on $P_{\max}$ to the e2e OPs:}
In Fig. \ref{eq_OP_vs_Pmax}, we compare the e2e OP of $\edgeuser$ and $\centeruser$ with the proposed ADM for UAV-aided coordinated DUL systems with different target spectral efficiency pairs; i.e., $\mathbf{R}_1 = (2,0.1)$ bits/s/Hz, $\mathbf{R}_2 = (1,0.05)$ bits/s/Hz, and $\mathbf{R}_3 = (0.5,0.025)$ bits/s/Hz, where ${ \mathbf{R}_j \triangleq ( R_{{\sf th},\centeruser}, R_{{\sf th},\edgeuser} ) }$.
    It is observed that the simulation results match the analytical results, which validate our analysis.

Under the proposed ADM, it is shown that ${\rm OP}_{\edgeuser \to \fusioncenter} $, ${\rm OP}^{[2]}_{\centeruser \to \fusioncenter} $, and ${\rm OP}^{[1]}_{\centeruser \to \fusioncenter}$ decrease as $P_{\max}$ increases where
\begin{align}
{\rm OP}_{\edgeuser,{\rm e2e}} 
    \ge {\rm OP}^{[2]}_{\centeruser,{\rm e2e}} 
    \ge {\rm OP}^{[1]}_{\centeruser,{\rm e2e}}.
\label{eq_OP_relation}
\end{align}

This can be explained as follows. 
Due to imperfect SIC, the decoding of $x_{\edgeuser}$ involves co-channel interference (CCI) terms ${ (1-\theta_1) P_{\max} g_{\edgeuser\uav}^2 {\pl}_{\edgeuser\uav} }$ and ${ \theta_1 P_{\max} | \tilde{h}_{\centeruser\uav} |^2 }$, and that of $\hat{x}_{\edgeuser}$ involves interference power of ${ \theta_1 P_{\max} g_{\centeruser\uav}^2 {\pl}_{\centeruser\uav} }$.
Moreover, the decoding of $x^{[2]}_{\centeruser}$ suffers from the interference from ${ (1-\theta_2) P_{\max} g_{\uav\fusioncenter}^2 {\pl}_{\uav\fusioncenter} }$ and the RI ${ (1-\theta_2) P_{\max} | {\tilde{h}}_{\uav\fusioncenter} |^2 }$.
    Meanwhile, $\fusioncenter$ can directly decode ${ x^{[1]}_{\centeruser} }$ without being affected by the CCIs, which results in the lowest e2e OP as in \eqref{eq_OP_relation}. 
Another observation from Fig. \ref{eq_OP_vs_Pmax} is that the outage performance can be improved by reducing $R_{{\sf th},\centeruser}$ and/or $R_{{\sf th},\edgeuser}$.
    Specifically, by halving both $R_{{\sf th},\centeruser}$ and $R_{{\sf th},\edgeuser}$, from 2.0 bits/s/Hz to 1.0 bits/s/Hz and from 0.1 bits/s/Hz to 0.05 bits/s/Hz, respectively, ${\rm OP}_{\edgeuser,{\rm e2e}} $, ${\rm OP}^{[1]}_{\centeruser,{\rm e2e}}$, and ${\rm OP}^{[2]}_{\centeruser,{\rm e2e}}$ at $P_{\max} = 35$ dBm decrease from $10^{-2.8}$, $10^{-1.8}$, and $10^{-0.5}$ to $10^{-3.9}$, $10^{-1.95}$, and $10^{-0.8}$, respectively.
In this case, the QoS requirements become less demanding, thus decreasing the e2e OPs. In other words, the e2e OPs are increasing functions of the target spectral efficiency, i.e., $R_{{\sf th}, \edgeuser}$ and $R_{{\sf th}, \centeruser}$.

\textit{Impact of $R_{{\sf th},\centeruser}$ and $R_{{\sf th},\edgeuser}$ on the system throughput:}
In Fig. \ref{fig_throughput_vs_specEff}, we illustrate the throughput versus $R_{{\sf th},\centeruser}$ and $R_{{\sf th},\edgeuser}$, when $\gamma$ is set to $20$. 
    It is observed that as $R_{{\sf th},\centeruser}$ increases above $0$ bits/s/Hz, the throughput gradually increases until reaching peak values.
For instance, the peak value of throughput when $\uav$ is located at the altitude $Z_{\uav} = 6.67$ [m] is 2.75 bits/s/Hz.
Beyond that value, the throughput drastically drops to $0$ bits/s/Hz.
    The reason for this is because while the proposed ADM can satisfy a wide variety of QoS requirements, they cannot meet higher demands of ground users, which leads to a drastic decrease in the system throughput.
    
\textit{Comparison with CDRT-OMA:}
In Fig. \ref{fig:3p1_1}, we compare the proposed CDRT-NOMA versus CDRT-OMA transmission.
    It is observed from the result in Fig. \ref{fig:3p1_1} that CDRT-NOMA is indeed superior than CDRT-OMA.
More importantly, as the target SE increases from $\mathbf{R}_3$ to $\mathbf{R}_1$, the gap in the throughput between CDRT-NOMA and CDRT-OMA increases, from 0.256 bits/s/Hz to 0.568 bits/s/Hz.
    The reason for such superiority is that the proposed CDRT-NOMA uses just two time slots for the full transmission, whereas the CDRT-OMA requires twice as many time slots to complete the same task.

\textit{Numerical Gradient Descend-based Algorithm:}
In Figs.~\ref{fig_NGD_convergence} and \ref{fig_BFS_NGD_vs_location}, we examine the accuracy of the obtained optimal power allocation coefficients using the proposed NGD-based algorithm. As can be seen in Fig.~\ref{fig_NGD_convergence}, the NGD-based algorithm converges to the optimal point. The step $\tau$ has strong impact on the convergence speed, the higher $\tau$, the faster convergence the algorithm can reach. In Fig.\ref{fig_BFS_NGD_vs_location}, we show that the proposed NGD-based algorithm results in similar optimal throughput as that achieved by the BFS algorithm, and the optimality gap between the two algorithms is relatively small. 
Specifically, let us define the mean squared error (MSE) as
$
{\rm MSE} \triangleq
\frac{1}{N}
\sum_{n = 1}^{N}
\left| 
    {\cal R}^{\ast,{\rm BFS}}_{\Sigma}[n]
    -   {\cal R}^{\ast,{\rm NGD}}_{\Sigma}[n]
\right|^2,
$
where $N$ denotes the number of data points,
    ${\cal R}^{\ast,{\rm BFS}}_{\Sigma}[n]$, and ${\cal R}^{\ast,{\rm NGD}}_{\Sigma}[n]$ are the optimal and sub-optimal system throughput obtained from the BFS and the NGD at the $n$-th data point, respectively. 
It can be observed that the proposed algorithm matches well with the BFS algorithm, with an MSE of $4.2583 \times 10^{-8}$ over $N = 10^2$ data points.

\textit{Impact of $\theta_1$ and $\theta_2$ on the system throughput:}
In Fig. \ref{fig:7}, we examine the throughput versus the power allocation in the first and the second phase, i.e., $\theta_1$ and $\theta_2$, respectively.
    The optimal value is obtained by using the brute-force search algorithm.
In addition, it can be observed that as $\theta_1$ and $\theta_2$ increase from zero to one, more power is allocated to both $x^{[1]}_{\centeruser}$ and $x^{[2]}_{\centeruser}$, and the throughput increases to the optimal value, which is marked by the red star.
    Beyond the optimal value, less power is allocated to $x_{\edgeuser}$ and $\hat{x}_{\edgeuser}$, thus causing a slight drop in the system throughput.

\textit{Impact of the residual interference level:}
In Fig. \ref{fig_throughput_vs_RILevel}, 
    we study the impact of the RI level on the system throughput of the proposed UAV-aided double uplink system. 
It can be observed that when $\xi$ increases from $-30$ dB to $0$ dB, the throughput decreases for both ADM and NADMs. 
    However, the throughput does not necessarily reach zero when $\xi$ is large, e.g., $\xi = 0$ dB.
This is because the decoding of ${ x_{\edgeuser} }$, $x^{[2]}_{\centeruser}$, and ${\hat{x}_{\edgeuser}}$ are affected by the RI power from $P^{[1]}_{\centeruser} |\tilde{h}_{\centeruser\uav}|^2$, 
    $P^{[2]}_{\centeruser} |\tilde{h}_{\centeruser\fusioncenter}|^2$, and
    $P_{\uav} |\tilde{h}_{\uav\fusioncenter}|^2$. Thus, as $\xi$ increases, the RI power increases, reducing the system throughput. 
However, the decoding of ${ x^{[2]}_{\centeruser} }$ is not affected by the RI power, thus resulting in a lower bound for the system throughput, such that
\begin{align}
\mathcal{R}_{\Sigma} (\xi)
    \ge \frac{ {R}_{\sf th,C} }{ 2 }
        (1-{\rm OP}^{[1]}_{\centeruser,{\rm e2e}}),~\forall \xi \in [0,1],
\end{align}
which ensures the proposed system also provides an acceptable throughput, even in the presence of high RI powers.

\textit{Accuracy of \eqref{eq_avgRSigma_pos}:}
In Fig. \ref{eq_OP_vs_angle}, we compare the average throughput in \eqref{eq_avgRSigma_pos} and in \eqref{eq:R_Sigma_GC}. 
    When the number of observations is insufficient, the gap between \eqref{eq_avgRSigma_pos} and \eqref{eq:R_Sigma_GC} is large, which indicates that \eqref{eq:R_Sigma_GC} shows low accuracy in low values of $N$.
As $N$ increases, the gap between \eqref{eq_avgRSigma_pos} and \eqref{eq:R_Sigma_GC} decreases until $N$ exceeds $10^3$, at which \eqref{eq:R_Sigma_GC} remains constant.
    Note that \eqref{eq_avgRSigma_pos} is an approximation of \eqref{eq:R_Sigma_GC} when $N \to \infty$, thus from Fig. \ref{eq_OP_vs_angle}, it is expected that the joint PDF of $d_{\uav\fusioncenter}$, $d_{\edgeuser\uav}$, $d_{\centeruser\fusioncenter}$, and $d_{\centeruser\uav}$, denoted as, $f_{d_{\uav\fusioncenter},d_{\edgeuser\uav},d_{\centeruser\fusioncenter},d_{\centeruser\uav}}(x,y,z,t)$ is tightly lower-bounded by
    $f_{d_{\uav\fusioncenter}}(x)    
    f_{d_{\edgeuser\uav}}(y)    
    f_{d_{\centeruser\fusioncenter}}(z)    
    f_{d_{\centeruser\uav}}(t)$.
    
\textit{Impact of locations and velocity:}
It is observed in Fig. \ref{fig_insRSigma_vs_Location} that the proposed ADM provides a stable system throughput, which maintains ${\cal R}_{\Sigma}$ around 0.51 bits/s/Hz in both scenarios.
    In contrast, when NADM is applied in scenarios involving fast-moving nodes, as seen in Figure \ref{fig_insRSigma_vs_Locationb}, the system throughput wildly fluctuates and becomes unpredictable.
    For instance, the conventional NADM with decoding order ${\bf d}_{\edgeuser\edgeuser}$ (i.e., the cyan line) causes the throughput to fluctuate between 0.25 bit/s/Hz to 0.48 bits/s/Hz. 
This is because when $\uav$ moves, the path losses between the nodes (i.e., ${\pl}_{\centeruser\uav}$, ${\pl}_{\edgeuser\uav}$, ${\pl}_{\uav\fusioncenter}$, and ${\pl}_{\centeruser\fusioncenter}$) also fluctuates rapidly. 
    In this context, NADM cannot enable $\uav$ and $\fusioncenter$ to adaptively adjust the decoding order according to this fluctuation, thus resulting in unpredictable outage behaviors. 
Since the throughput, in this paper, is based on the e2e OPs, the throughput obtained using NADM also fluctuates significantly.
    In addition, it is observed in Fig. \ref{fig_insRSigma_vs_Locationa} that NADM-aided CDRT-NOMA can stabilize the throughput around 0.5 [bits/s/Hz] when nodes moves slowly.

\vspace{-5pt}
\section{Conclusion}
In this paper, we proposed ADM for UAV-aided DUL CDRT-NOMA networks. 
    Specifically, by exploiting the channel gains from the transmitters (i.e., the terrestrial users and the $\uav$), the proposed ADM decide on the decoding order at the receivers in an adaptive manner.
To evaluate the performance of the proposed mechanisms, we derived exact closed-form expressions for the e2e OP and the system throughput. 
We showed that the proposed ADM can provide better performance in terms of OPs and throughput than the traditional NADMs.
    To study the impact of $\uav$'s trajectory, we modeled the locations of $\uav$ over time using the Random Waypoint Mobility (RWM) model, and we showed that the proposed ADM can maintain the system throughput regardless of the locations of $\uav$.
In contrast, the performance of the NADMs significantly fluctuate over the locations of $\uav$.
    To study the optimal throughput, we proposed a sub-optimal Gradient Descent-based algorithm and compared it to the BFS algorithm.
The results show that the proposed sub-optimal algorithm matches the BFS algorithm.
    In order to evaluate the sustainability of the proposed ADM in nonstationary scenarios, we assume that both the ground UEs and the UAV are mobile in accordance with the RWM model and the Reference Point Group Mobility (RPGM) model, respectively. 
    Additionally, we provided accurate formulas for each of the distance distributions. The numerical results reveal that the ADM-enhanced NOMA not only outperforms OMA, but also improves the performance of UAV-enabled UL-NOMA even in mobile environments.

\appendices
\vspace{-0.2cm}
\section{Proof of Lemma \ref{lemma_realValued_Gamma}} \label{apx_realValued_Gamma}

The Laplace transformation of $g$ is obtained as $\mathfrak{L}_{g}(s) \!=\! \frac{1}{(1+\frac{s}{m})^m}$.
For $\varepsilon \!\triangleq\! \lfloor m \rfloor \!-\! m$, i.e., $\varepsilon$ is the real part of $m$, the term $\left( 1+\frac{s}{m} \right)^{ -\varepsilon }$ can be approximated as
\begin{equation}
\left(
	1+\frac{s}{m}
\right)^{ -\varepsilon } 
\approx
	\sum_{k=1}^{M} \frac{ w_k }{ 1+\sigma_k s },~s > 0,
\label{eq_approx_gamma_apx}
\end{equation}
where $0 \!\le\! w_t \!\le\! 1$ satisfying $\sum_{t=1}^{M}{ w_t } \!=\! 1$ and $\sigma_k > 0$ are coefficients obtained from fitting the left-hand side of \eqref{eq_approx_gamma_apx} to the right-hand side of \eqref{eq_approx_gamma_apx} using curve fitting process.

By applying partial fraction decomposition, $\mathfrak{L}_{g}(s)$ can be accurately approximated as
\begin{align}
\mathfrak{L}_{g}(s)
    &\approxeq
    \sum_{k=1}^M w_k
    \frac{ m^m }{ \sigma_k }
    \bigg\{
    	\sum_{t=1}^{ \lfloor m \rfloor }
        \frac{ R_{1}(t,k) }{ (s+m)^{t} }
        + \frac{ R_{2}(k) }{ s+\sigma_k^{-1} }
    \bigg\},
\label{eq_Laplace_G_partial}\\
    &\approxeq
	\sum_{t=1}^{m} \upzeta_{t,M}
		\left( 1+\frac{s}{m} \right)^{-t}
	+ \sum_{k=1}^{M} \frac{ \upzeta_{k} }{1+\sigma_k s}
,~ s > 0,
\label{eq_Laplace_G_final}
\end{align}
where $R_{1,k} \triangleq -\frac{1}{( m-\theta_k^{-1} )^{\lfloor m \rfloor-t+1}}$ and 
    $R_{2,k} \triangleq \frac{1}{( m-\theta_k^{-1})^{\lfloor m \rfloor}}$, for $k \ge 1$. 
By applying inverse Laplace transform onto \eqref{eq_Laplace_G_final}, i.e., $f_{g}(x) = \mathfrak{L}^{-1}\{ \mathfrak{L}_{g}(s);s,x \}$, we then obtain \eqref{eq:PDF_Gamma_Aprx_fin}. This completes the proof.

\vspace{-0.3cm}
\section{Proof of \eqref{eq:Wn_CFE}} \label{apx_lemma_2}

By applying \cite[Eq. (8.352.2)]{Gradshteyn2007}, where
${\Gamma(n,x) = \Gamma(n) e^{-x} \sum_{k=0}^{n-1} { \frac{x^k}{k!} }}$, $\mathpzc{W}_1({\bf W}_1)$ can be obtained~as
\begin{align}
\mathpzc{W}_1({\bf W}_1) =
	\sum_{k_1 = 1}^{ \mathclap{K_1 + 1} }
		\chi_1(k_1) 
		e^{	- \frac{p_1 x_2 + q_1}{\Lambda_1} }
	\sum_{i_1=0}^{\kappa_1-1}
		\frac{\big(
			\frac{p_1 x_2 + q_1}{\Lambda_1}
		\big)^{i_1}}{i_1!}.
\end{align}
Subsequently, by applying the binomial theorem, we get
\begin{align}
\left(\frac{p_1 x_2 + q_1}{\Lambda_1}\right)^{i_1} 
	= \frac{ i_1! }{(\Lambda_1)^{i_1}}
	\sum_{j_1+l_1=i_1} \frac{(p_1 x_2)^{l_1}}{l_1!} \frac{(q_1)^{j_1}}{j_1!}
\nonumber
\end{align} 
and utilizing \eqref{eq_pdf_G_final}, the integral $\mathpzc{W}_1({\bf W}_1)$ can be obtained as
\begin{align}
&\mathpzc{W}_2({\bf W}_2) = \sum_{k_1 = 1}^{\mathclap{K_1 + 1}}
		\Xi_1
	\sum_{i_1=0}^{\kappa_1-1}
		\Lambda_1^{-i_1} 
	\sum_{ \mathclap{j_1+l_1=i_1} } \quad
		\frac{ e^{- \frac{q_1}{\Lambda_1} } }
		{ (\Lambda_1)^{-\kappa_1} } 
		\Gamma(\kappa_1)
		\frac{(p_1)^{l_1}}{l_1!}
\nonumber \\
&\quad \times \frac{(q_1)^{j_1}}{j_1!} 
	\quad \sum_{k_2=1}^{ \mathclap{K_2+1} }
		\Xi_2
	\int_{p_2 x_3 + q_2}^\infty
	 	{x_2}^{\kappa_2-1} 
	 	e^{ -\frac{x_2}{\Lambda_2} } {\rm d} x_2,
\end{align}
which can be derived by repeating similar steps when deriving $\mathpzc{W}_1({\bf W}_1)$ and $\mathpzc{W}_1({\bf W}_1)$.
By induction, we then obtain $\mathpzc{W}_n({\bf W}_n)$ after repeating the above steps. This completes the proof of \eqref{eq:Wn_CFE}.
\vspace{-0.4cm}

\section{Proof of Lemma \ref{lem_closedForm_P1}} \label{apx:B}
The first probability in \eqref{eq_Pout_uav_sim} can be expressed as
\begin{align} 
P^{[\uav]}_{\centeruser\to\edgeuser}
&= \Pr\left[
    \phi_{\centeruser\uav} \!>\! \alpha_1 \phi_{\edgeuser\uav} + a_1,
    \phi_{\edgeuser\uav} \!>\! \alpha_2 \varphi_{\centeruser\uav} + a_2, \phi_{\centeruser\uav} \!>\! \phi_{\edgeuser\uav}
\right] \nonumber \\
&= \Pr\left[
	\phi_{\centeruser\uav} \!>\! \alpha_1 \phi_{\edgeuser\uav} + a_1 >
    \phi_{\edgeuser\uav} \!>\! \alpha_2 {\varphi}_{\centeruser\uav} \!+\! a_2
\right] \label{eq_80} \\
&\quad\!+\! 
\Pr\left[
	\phi_{\centeruser\uav} > \phi_{\edgeuser\uav} > \alpha_2 \varphi_{\centeruser\uav} + a_2, 
    \alpha_1 \phi_{\edgeuser\uav} + a_1 < \phi_{\edgeuser\uav}
\right].\nonumber
\end{align}
    For convenience, we denote the first and the second probability of \eqref{eq_80} as $\rho_{1}$ and $\rho_{2}$, respectively.
\vspace{-5pt}
\subsection{The probability $\rho_{1}$}

We consider two subcases:

\noindent + \textit{Subcase 1}: $\alpha_1 < 1$ and $A_1 > a_2$, thus
\begin{align}
{\rho}_{1} 
&=
\Pr\big[
	\phi_{\centeruser\uav} \ge \alpha_1 \phi_{\edgeuser\uav} + a_1,
\nonumber\\
&\qquad\qquad
	A_1 > \phi_{\edgeuser\uav} \ge \alpha_2 \varphi_{\centeruser\uav} \!+\! a_2,
	\mathcal{A} > \varphi_{\centeruser\uav}
\big],
\nonumber\\
&\mathop{=}\limits^{(a)}  
\Pr\left[
	\phi_{\centeruser\uav} \ge \alpha_1 \phi_{\edgeuser\uav} + a_1, \phi_{\edgeuser\uav} > \alpha_2 \varphi_{\centeruser\uav} + a_2,
	\varphi_{\centeruser\uav} < {\cal A}
\right] 
\nonumber\\
&\quad- \Pr\left[
	\phi_{\centeruser\uav} \ge \alpha_1 \phi_{\edgeuser\uav} + a_1, \phi_{\edgeuser\uav} > A_1, \varphi_{\centeruser\uav} < {\cal A}
\right]
\nonumber\\
&\triangleq \rho_{1,1} - \rho_{1,2},
\end{align}
where $(a)$ is due to $\Pr[E_a E_b] = \Pr[E_a]-\Pr[E_a\overline{E}_b]$ for two random events $E_a$ and $E_b$. 

By adopting \eqref{eq:Wn_CFE} and \eqref{eq:wn_CFE}, we can express $\rho_{1,1}$ and $\rho_{1,2}$ as
\begin{align}
\rho_{1,1} 
    &=  \mathpzc{W}_3(
        \hat{\Theta}_{\centeruser\uav}, \Theta_{\edgeuser\uav},  \Theta_{\centeruser\uav}, 0, \alpha_2, \alpha_1,
        {\cal A}, a_2, a_1), \nonumber\\
\rho_{1,2} 
    &=  \mathpzc{w}_3(
        \hat{\Theta}_{\centeruser\uav}, \Theta_{\edgeuser\uav}, \Theta_{\centeruser\uav}, 0, 0, \alpha_1, {\cal A}, A_1, a_1)
\nonumber\\
    &=  \mathpzc{w}_1(\hat{\Theta}_{\centeruser\uav}, 0, {\cal A})
        \mathpzc{W}_2(\Theta_{\edgeuser\uav}, \Theta_{\centeruser\uav},
        0, \alpha_1, A_1, a_1).
\nonumber
\end{align}

\noindent + \textit{Subcase 2}: $\alpha_1 \ge 1$, thus
\begin{equation}
{\rho}_{1} 
    = \Pr\left[
	\phi_{\centeruser\uav} \ge \alpha_1 \phi_{\edgeuser\uav} + a_1, \phi_{\edgeuser\uav} > \alpha_2 \varphi_{\centeruser\uav} + a_2
\right] = \rho_{1,2}.
\end{equation}
By combining the results of the two subcases, we obtain that
\begin{equation}
{\rho}_{1} 
=  \left\{
\begin{array}{cl}
    \rho_{1,1} - \rho_{1,2},
	&  {\rm if}~\alpha_1 < 1, A_1 > a_2, \\
	\rho_{1,2},
	& {\rm if}~\alpha_1 \ge 1.
\end{array} \right.
\label{eq_vrho_1}
\end{equation}

\vspace{-0.3cm}
\subsection{The probability $\rho_{2}$}

We find that ${\rho}_{2} \!=\! 0$ if $\alpha_1 \!\ge\! 1$. 
    Considering the case $\alpha_1 < 1$, we have
{\allowdisplaybreaks
\begin{align}
{\rho}_{2} 
&= \Pr\left[
	\phi_{\centeruser\uav} \ge \phi_{\edgeuser\uav} \ge \alpha_2 \varphi_{\centeruser\uav} + a_2, \phi_{\edgeuser\uav} > A_1
\right] \nonumber\\
&= \Pr\left[
	\phi_{\centeruser\uav} \!\ge\! \phi_{\edgeuser\uav} \!\ge\! \alpha_2 \varphi_{\centeruser\uav} + a_2, 
	\alpha_2 \varphi_{\centeruser\uav} + a_2 \ge A_1
\right] \nonumber \\
&\quad + \Pr\left[
	\phi_{\centeruser\uav} \ge \phi_{\edgeuser\uav} \ge A_1, \alpha_2 \varphi_{\centeruser\uav} + a_2 < A_1 
\right].
\end{align}}

Next, we consider two subcases:

\noindent + \textit{Subcase 1}: $A_1 > a_2$, thus
{\allowdisplaybreaks
\begin{align}
{\rho}_{2} 
&= \Pr\left[
	\phi_{\centeruser\uav} \ge \phi_{\edgeuser\uav} \ge \alpha_2 \varphi_{\centeruser\uav} + a_2, 
	\varphi_{\centeruser\uav} > \mathcal{A}
\right] \nonumber \\
&\quad+ \Pr\left[
	\phi_{\centeruser\uav} \ge \phi_{\edgeuser\uav} \ge A_1, \varphi_{\centeruser\uav} < \mathcal{A}
\right] \nonumber \\
&= \Pr\left[
	\phi_{\centeruser\uav} \ge \phi_{\edgeuser\uav} \ge \alpha_2 \varphi_{\centeruser\uav} + a_2, 
	\varphi_{\centeruser\uav} > \mathcal{A}
\right] \nonumber \\
&\quad
- \Pr\left[
	\phi_{\centeruser\uav} \ge \phi_{\edgeuser\uav} \ge A_1
\right] \Pr[\varphi_{\centeruser\uav} < \mathcal{A}]
\triangleq
    \rho_{2,1} - \rho_{2,2},
\end{align}}
where $\rho_{2,1}$ and $\rho_{2,2}$ are obtained as
\begin{subequations}
\begin{align}
\rho_{2,1} 
    &=  \mathpzc{W}_3(
        \hat{\Theta}_{\centeruser\uav}, \Theta_{\edgeuser\uav}, \Theta_{\centeruser\uav}, 0, \alpha_2, 1, {\cal A}, a_2, 0), \\
\rho_{2,2} 
    &=  \mathpzc{W}_1(\hat{\Theta}_{\centeruser\uav}, 0, {\cal A})
    \mathpzc{W}_2(\Theta_{\edgeuser\uav}, \Theta_{\centeruser\uav},
        0, 1, A_1, 0),
\end{align}
\end{subequations}
    
\noindent + \textit{Subcase 2}: $A_1 \le a_2$, thus
\begin{align}
{\rho}_{2} 
&=   \Pr\left[
	\phi_{\centeruser\uav} > \phi_{\edgeuser\uav} > \alpha_2 \varphi_{\centeruser\uav} + a_2
\right] \\
&=  \mathpzc{W}_3(\hat{\Theta}_{\centeruser\uav}, \Theta_{\edgeuser\uav},  \Theta_{\centeruser\uav}, 0, \alpha_2, 1, 0, a_2, 0)
\triangleq {\rho}_{2,3}.
\end{align}
By combining the results of the two subcases, we obtain that
\begin{align}
\rho_2 
    =   \left\{
    \begin{array}{cl}
        0 &, \alpha_1 \ge 1,  \\
        \rho_{2,1} - \rho_{2,2} &, \alpha_1 < 1, A_1 > a_2 \\
        \rho_{2,3}&, \alpha_1 < 1, A_1 > a_2
    \end{array}
    \right.
\label{eq_vrho_2}
\end{align}
By substituting \eqref{eq_vrho_1} and \eqref{eq_vrho_2} into \eqref{eq_80}, we obtain \eqref{eq:Lem3_final}. This completes the proof of Lemma \ref{lem_closedForm_P1}.
\vspace{-0.4cm}
\section{Proof of Lemma \ref{lem:2}} \label{apx:C}
The second probability in \eqref{eq_Pout_uav_sim} can be expressed as
\begin{align}
P^{[\uav]}_{\edgeuser \to \centeruser}
&=  \Pr\left[
	\phi_{\edgeuser\uav} > \alpha_2 \phi_{\centeruser\uav} + a_2
    \phi_{\edgeuser\uav} < \phi_{\centeruser\uav}
\right] 
\nonumber \\
&= \Pr\left[
	\phi_{\edgeuser\uav} > \alpha_2 \phi_{\centeruser\uav} + a_2,
	a_2 + \alpha_2 \phi_{\centeruser\uav} > \phi_{\centeruser\uav}
\right] \nonumber\\
&\quad + \Pr\left[
	\phi_{\edgeuser\uav} > \phi_{\centeruser\uav}, 
	a_2 + \alpha_2 \phi_{\centeruser\uav} < \phi_{\centeruser\uav}
\right].
\end{align}
In the case $\alpha_2 < 1$, we can rewrite $P^{[\uav]}_{\edgeuser \to \centeruser}$ as
\begin{align}
P^{[\uav]}_{\edgeuser \to \centeruser}
&=  \Pr\left[
	\phi_{\edgeuser\uav} > \alpha_2 \phi_{\centeruser\uav} + a_2,
    \phi_{\centeruser\uav} > A_1
\right] \\
&\quad
- \Pr\left[
	\phi_{\edgeuser\uav} > \alpha_2 \phi_{\centeruser\uav} + a_2
\right]
+ \Pr\left[
	\phi_{\edgeuser\uav} > \phi_{\centeruser\uav} > A_1
\right]. \nonumber
\end{align}

Let us denote the first, the second, and the third probability as
$\rho_{3}$, $\rho_{4}$, and $\rho_{5}$, respectively, by adopting \eqref{eq_In_def_prob},
    we can express $\rho_{3}$, $\rho_{4}$ and $\rho_{5}$ as
\begin{align}
\rho_{3} &=  
\mathpzc{W}_1(\Theta_{\centeruser\uav}, \Theta_{\edgeuser\uav}, 
    \alpha_2, a_2, 0, A_1), \\
\rho_{4} &=  
\mathpzc{W}_1(\Theta_{\centeruser\uav}, \Theta_{\edgeuser\uav},
    \alpha_2, a_2, 0, 0), \\
\rho_{5} &=  
\mathpzc{W}_1(\Theta_{\centeruser\uav}, \Theta_{\edgeuser\uav},
    1, 0, 0, A_1).
\end{align}
In the case $\alpha_2 \ge 1$, we have $P^{[\uav]}_{\edgeuser \to \centeruser} = \rho_{4}$. Combining the foregoing results, we obtain \eqref{eq:Lem4_final}. This completes the proof of Lemma \ref{lem:2}.

\vspace{-0.4cm}
\section{Proof of Lemma \ref{lem:dEF}} \label{apx:D}
The joint PDF of $d_{\edgeuser,0}$ and $\varphi_{\edgeuser,0}$ can be obtained as \cite{HyytiaTMC2006}
\begin{align}
f_{d_{\edgeuser,0},\varphi_{\edgeuser,0}}(r,\varphi) 
    =   r \frac{\eta(R;r)}{C_\eta(R)}
    \approxeq r \frac{P_3(R;r)}{C_P(R)},
\end{align}
for $r \in (0,R)$ and $\varphi \in (0,2 \pi)$.

Since $d_{\edgeuser,0}\!=\!\sqrt{(X_{\edgeuser}-X_{\edgeuser,0})^2+(Y_{\edgeuser}-Y_{\edgeuser,0})^2}$ when 
    $Z_{\edgeuser} = 0$, we have 
    $X_{\edgeuser}-X_{\edgeuser,0} \!=\! d_{\edgeuser,0} \cos\varphi_{\edgeuser,0}$ and
    $Y_{\edgeuser}-Y_{\edgeuser,0} \!=\! d_{\edgeuser,0} \sin\varphi_{\edgeuser,0}$. The joint PDF of ${X_{\edgeuser}-X_{\edgeuser,0}}$ and ${Y_{\edgeuser}-Y_{\edgeuser,0}}$ is obtained~as
$f_{X_{\edgeuser}-X_{\edgeuser,0},Y_{\edgeuser}-Y_{\edgeuser,0}}(x,y)
    \!=\!   f_{d_{\edgeuser,0},\varphi_{\edgeuser,0}}(\sqrt{x^2+y^2},\varphi)
$, for $x^2 + y^2 < R^2$. Hence, the joint PDF of ${X_{\edgeuser}}$ and ${Y_{\edgeuser}}$ is obtained~as
\begin{align}
f_{X_{\edgeuser},Y_{\edgeuser}}(x,y)
    =   f_{X_{\edgeuser},Y_{\edgeuser}}(x+X_{\edgeuser,0},y+Y_{\edgeuser,0}),
\end{align}
for $(x+X_{\edgeuser,0})^2 + (y+Y_{\edgeuser,0})^2 < R^2$. 

Note that ${X_{\edgeuser} = d_{\edgeuser\fusioncenter} \cos\varphi_{\edgeuser\fusioncenter}}$ and ${Y_{\edgeuser} = d_{\edgeuser\fusioncenter}\sin\varphi_{\edgeuser\fusioncenter}}$, the joint PDF of $d_{\edgeuser\fusioncenter}$ and $\varphi_{\edgeuser\fusioncenter}$ is obtained as
\begin{align}
f_{d_{\edgeuser\fusioncenter},\varphi_{\edgeuser\fusioncenter}}(r,\varphi)
    =   r f_{X_{\edgeuser},Y_{\edgeuser}}(r\cos\varphi,r\sin\varphi),
\end{align}
for $\max(D_0-R,0) < r < D_0+R$. Herein, $\varphi \in (0,2\pi)$ and satisfies $(r\cos\varphi+X_{\edgeuser,0})^2 + (r\sin\varphi+Y_{\edgeuser,0})^2 < R^2$. Hence, the PDF of $d_{\edgeuser\fusioncenter}$ is obtained as
\begin{align}
f_{d_{\edgeuser\fusioncenter}}(r)
    =   \int_{C(\varphi)}
    f_{d_{\edgeuser\fusioncenter},\varphi_{\edgeuser\fusioncenter}}(r,\varphi) {\rm d} \varphi,
\label{eq_PDF_dEF_integral}
\end{align}
where ${C(\varphi) \triangleq \cos(\varphi - \varphi_{\edgeuser,0}) \!<\! \frac{r^2+D_0^2-R^2}{2 r D_0}}$.
By solving the inequality 
    $C(\varphi)$ for $\varphi$, \eqref{eq_PDF_dEF_integral} can be derived via the two following cases: 
\begin{itemize}
    \item[+] Case $\frac{r^2+D_0^2-R^2}{2 r D_0} > 1$:     
    We obtain $r \!<\! R \!-\! D_0$ when $R \!>\! D_0$ and $C(\varphi)$ accepts all values of $\varphi$ in the range $[0,2\pi]$. 
    \item[+] Case $-1 \!<\! \frac{r^2+D_0^2-R^2}{2 r D_0} \!<\! 1$:        
    We obtain ${R+D_0 > r > \max(R-D_0,0)}$ and $C(\varphi)$ accepts $\varphi$ in the range from ${\varphi_{\edgeuser,0}-\nu_c(r)}$ to ${\varphi_{\edgeuser,0}+\nu_c(r)}$, where $\nu_c = \cos^{-1} \big( \frac{r^2+D_0^2-R^2}{2 r D_0} \big)$.
\end{itemize}
For both cases, we adopt the binomial theorem, the identities \cite[Eq. (2.631.7)]{Gradshteyn2007} and \cite[Eq. (2.631.2)]{Gradshteyn2007} to obtain the closed-form expressions of $f_{d_{\edgeuser\fusioncenter}}(r)$. 
The step-by-step details of the derivation are quite tedious, thus we omit them from this proof. This completes the proof of Lemma \ref{lem:dEF}.

\vspace{-0.2cm}
\bibliographystyle{IEEEtran}
\bibliography{Refs_UL_NOMA}

\end{document}